\documentclass{cai}

\usepackage{graphics}
\usepackage{epstopdf}
\usepackage{epsfig}

\usepackage{color}
\usepackage{epic,eepic}

\usepackage{amsfonts}
\usepackage{latexsym}
\usepackage{amssymb}
\usepackage{amsmath}

\usepackage{float}
\usepackage{xspace}

\usepackage{hyperref}

\usepackage{graphics}
\usepackage{epstopdf}
\usepackage{epsfig}

\newcommand{\umod}{{\bf~\%~}\xspace}
\newcommand{\uassign}{\leftarrow}
\newcommand{\uif}{{\bf if}\xspace}
\newcommand{\uthen}{{\bf then}\xspace}
\newcommand{\uelse}{{\bf else}\xspace}

\newcommand{\urepeat}{{\bf repeat}\xspace}
\newcommand{\uuntil}{{\bf until}\xspace}
\newcommand{\ufor}{{\bf for}\xspace}
\newcommand{\uto}{{\bf to}\xspace}

\newcommand{\udo}{{\bf do}\xspace}
\newcommand{\ureturn}{{\bf return}\xspace}
\newcommand{\ubreak}{{\bf break}\xspace}
\newcommand{\ucontinue}{{\bf continue}\xspace}

\newcounter{lineno}

\newcommand{\utab}{\qquad}

\newcommand{\startindent}{\hspace{0.8em}}

\newenvironment{code}{%
\setcounter{lineno}{0}%
\begin{tabbing}
\utab\=\utab\=\utab\=\utab\=\utab\=\utab\=\utab\=%
\utab\=\utab\=\utab\=\utab\=\utab\=\utab\= \kill
}
{
\end{tabbing}%
\vspace{-2mm}
}

\floatstyle{ruled}
\newfloat{algorithm}{tbp}{flt}
\floatname{algorithm}{\small\bf Alg.}

\newcommand{\floor}[1]{{\lfloor #1 \rfloor}}

\newcommand{\exclude}[1]{}

\def\mmod{~\textrm{mod}~}

\begin{document}
\label{firstpage}

\title[Two simple full-text indexes based on the suffix array]
      {Two Simple Full-Text Indexes\\ based on the Suffix Array}

\author[Sz.~Grabowski, M.~Raniszewski]
       {Szymon \surname{Grabowski}, Marcin \surname{Raniszewski}}

\affiliation{Institute of Applied Computer Science\\
Lodz University of Technology\\
Al.\ Politechniki 11\\
90--924 {\L}\'od\'z, Poland}

\email{sgrabow@kis.p.lodz.pl, mranisz@kis.p.lodz.pl}

%
%

\noreceived{} \nocommunicated{}

\maketitle

\begin{abstract}
We propose two suffix array inspired full-text indexes.
One, called SA-hash, augments the suffix array with a hash table 
to speed up pattern searches due to significantly narrowed search 
interval before the binary search phase.
The other, called FBCSA, is a compact data structure, similar to
M{\"a}kinen's compact suffix array, but working on fixed sized blocks.
Experiments 
on the Pizza~\&~Chili 200\,MB datasets 
show that SA-hash is about 2--3 times faster in pattern searches (counts) 
than the standard suffix array, for the price of requiring $0.2n-1.1n$ 
bytes of extra space, where $n$ is the text length, and setting a 
minimum pattern length.
FBCSA is relatively fast in single cell accesses (a few times faster 
than related indexes at about the same or better compression), 
but not competitive if many consecutive cells are to be extracted. 
Still, for the task of extracting, e.g., 10 successive cells its time-space 
relation remains attractive.
\end{abstract}



\section{Introduction}

The field of text-oriented data structures continues to bloom.
Curiously, in many cases several years after ingenious theoretical solutions
their more practical (which means: faster and/or simpler) counterparts are presented, 
to mention only recent advances in rank/select implementations~\cite{GP14} 
or the FM-index reaching the compression ratio bounded by $k$-th order
entropy with very simple means~\cite{DBLP:conf/spire/KarkkainenP11}.

Despite the great interest in compact or compressed\footnote{%
By the latter we mean indexes with space use bounded by $O(n H_0)$ or even 
$O(n H_k)$ bits,  where $n$ is the text length,
and $H_0$ and $H_k$ respectively the order-0 and the order-$k$ entropy.
The former term, compact full-text indexes, is less definite, 
and, roughly speaking, may fit any structure with less than $n\log_2 n$ 
bits of space, at least for ``typical'' texts.}
full-text indexes in recent years~\cite{NMacmcs06}, we believe that in 
some applications search speed is more important than memory savings, thus 
different space-time tradeoffs are worth being explored.
The classic suffix array (SA)~\cite{MM90}, combining speed, simplicity and often 
reasonable memory use, may be a good starting point for such research.

In this paper we present two SA-based full-text indexes, combining effectiveness 
and simplicity.
One augments the standard SA with a hash table to speed up searches, 
for a moderate overhead in the memory use, 
the other is a byte-aligned variant of M{\"a}kinen's compact 
suffix array~\cite{DBLP:conf/cpm/Makinen00,DBLP:journals/fuin/Makinen03}.

A preliminary version of this article appeared in Proc. PSC 2014~\cite{GR14}.

\section{Preliminaries}
\label{sec:prelim}

We use 0-based sequence notation, that is, a sequence $S$ of length $n$ 
is written as $S[0 \ldots n-1]$, or equivalently as $s_0 s_1 \ldots s_{n-1}$.

One may define a {\em full-text index} over text $T$ of length $n$ 
as a data structure supporting at least two basic types of queries, 
both with respect to a pattern $P$ of length $m$, 
where $T$ and $P$ share an integer alphabet of size $\sigma$.
One query type is {\em count}: return the number $occ \geq 0$ of occurrences 
of $P$ in $T$.
The other query type is {\em locate}: for each pattern occurrence 
report its position in $T$, that is, such $j$ that $P[0 \ldots m-1] = T[j \ldots j+m-1]$.

The {\em suffix array} $SA[0 \dots n-1]$ for text $T$ is a permutation 
of the indexes $\{0, 1, \ldots, n-1\}$ such that 
$T[SA[i] \ldots n-1] \prec T[SA[i+1] \ldots n-1]$ for all $0 \leq i < n - 1$, 
where the ``$\prec$'' relation is the lexicographical order.
The inverse suffix array $SA^{-1}$ is the inverse permutation of $SA$, 
that is, $SA^{-1}[j] = i \Leftrightarrow SA[i] = j$.

If not stated otherwise, all logarithms throughout the paper are in base 2.

\section{Related work}
\label{sec:rwork}

The full-text indexing history starts with the {\em suffix tree} 
(ST)~\cite{Wei73}, a trie whose string collection is the set of all the 
suffixes of a given text, with an additional requirement that all 
non-branching paths of edges are converted into single edges. 

Typically, each ST path is truncated as soon as it points to a unique suffix. 
A leaf in the suffix tree holds a pointer to the text location where the 
corresponding suffix starts.
As there are $n$ leaves, no non-branching nodes and edge labels represented 
with pointers to the text, the suffix tree takes $O(n)$ words of space, 
which is in turn $O(n \log n)$ bits.

Although initially the suffix tree was known to be constructible in 
linear time only for constant alphabets, 
later an ingenious $O(n)$-time algorithm for integer 
alphabets was found~\cite{Far97}.
%
Moreover, the linear-time construction algorithms can 
be fast not only in theory, but also in practice~\cite{Gri07}.
Assuming constant-time access to any child of a given node, 
the search in the ST takes only $O(m + occ)$ time in the worst case.
In practice, this is cumbersome for a large alphabet, 
of size $n^{\omega(1)}$, 
as it requires using 
perfect hashing, which also makes the construction time linear only in expectation.
A small alphabet is easier to handle, which goes in line with the wide 
use of the suffix tree in bioinformatics.

The main problem with the suffix tree 
is its large space requirement.
Even in the most economical version~\cite{KB00} 
the ST space use 
reaches almost $9n$ bytes on average and $16n$ in the worst case, 
plus the text, for $\sigma \leq 256$, and even more for large alphabets.
Most implementations need $20n$ bytes or more.

An important alternative to the suffix tree is the {\em suffix array} (SA)~\cite{MM90}.
It is an array of $n$ pointers to all text suffixes 
sorted according to the lexicographic order of these suffixes.
The SA needs $n\log n$ bits for its $n$ suffix pointers (indexes), 
plus $n\log\sigma$ bits for the text, which typically translates to $5n$ 
bytes in total.
The pattern search time is $O(m\log n)$ in the worst case 
and $O(m\log_{\sigma}n + \log n)$ on average,
which can be
improved to $O(m + \log n)$ in the worst case 
using the longest common prefix (lcp) table.
Alternatively, the $O(m + \log n)$ time can be reached even without the lcp,
in a more theoretical solution with a specific suffix permutation~\cite{DBLP:journals/talg/FranceschiniG08}.
Yet Manber and Myers in their seminal paper~\cite{MM90} presented a nice 
trick saving several first steps in the binary search:
if we know the SA intervals for all the possible first $k$ symbols of the
pattern, we can immediately start the binary search in a corresponding interval.
We can set $k$ close to $\log_{\sigma} n$, with $O(n\log n)$ extra bits of
space, but constant expected size of the interval, which leads to $O(m)$ average
search time and only 
$O(\lceil m/CL \rceil)$ 
cache misses on average, 
where 
$CL$
is the cache line length expressed in symbols,  
typically 64 symbols/bytes in a modern CPU.
Unfortunately, real texts are far from random, hence in practice, 
if text symbols are bytes, 
we can use $k$ up to 3, which offers a limited (yet, non-negligible)
benefit.
This idea, later denoted as using a lookup table (LUT), is fairly well known, 
see e.g. its impact in the search over a suffix array on words~\cite{DBLP:conf/cpm/FerraginaF07}.

The suffix array can be built from the suffix tree by visiting its leaves 
in order (hence preserving $O(n)$ construction time), yet this approach is 
impractical.
Only in 2003 several algorithms building the SA directly in linear time 
were presented, e.g.,~\cite{KS03}, and currently the fastest $O(n)$-time 
construction algorithm is the one given by Nong~\cite{NZC11}.

A number of suffix tree or suffix array inspired indexes have been proposed 
as well, including the suffix cactus~\cite{karkkainen1995suffix} and the 
enhanced suffix array (ESA)~\cite{abouelhoda2002enhanced}, with space use usually 
between SA and ST, 
but according to our knowledge they generally are not faster than their 
famous predecessors in the count or locate queries.

On a theoretical front, the suffix tray by Cole et al.~\cite{cole2006suffix} 
allows to achieve $O(m + \log\sigma)$ search time, with $O(n)$ worst-case 
time construction and $O(n\log n)$ bits of space, which was recently improved 
by Fischer and Gawrychowski~\cite{FG15} 
to $O(m + \log\log\sigma)$ deterministic time, with preserved construction 
cost complexities.

The common wisdom about the practical performance of ST and SA 
is that they are comparable, but Grimsmo in his interesting experimental 
work~\cite{Gri07} showed that a careful ST implementation 
may be up to about 50\% faster than SA if the number of matches is very small 
(in particular, one hit), but if the number of hits grows, the SA becomes 
more competitive, sometimes being even about an order of magnitude faster.
Another conclusion from Grimsmo's experiments is that the ESA may also be 
moderately faster than SA if the alphabet is small (say, up to 8 symbols) 
but SA easily wins for a large alphabet.

Since around 2000 we can witness a great interest in succinct data structures, 
in particular, text indexes.
Two main ideas that deserve being mentioned are 
the compressed suffix array (CSA)~\cite{GV00,DBLP:conf/soda/Sadakane02} 
and the FM-index~\cite{FM00}; the reader is referred to the survey~\cite{NMacmcs06} 
for an 
thorough
coverage of the area.

It was noticed in extensive experimental comparisons~\cite{FGNVjea08,GP14} 
that compressed indexes are not much slower, and sometimes comparable, 
to the suffix array in count queries, but locate is 2--3 orders of magnitude 
slower if the number of matches is large.
This instigated researchers to follow one of two paths in order to 
mitigate the locate cost for succinct indexes.
One, pioneered by M{\"a}kinen~\cite{DBLP:conf/cpm/Makinen00,DBLP:journals/fuin/Makinen03} 
and addressed in a different way by Gonz{\'a}lez et al.~\cite{DBLP:conf/cpm/GonzalezN07,GNFjea14},
exploits repetitions in the suffix array (the idea is explained in Section~\ref{sec:fbsa}).
The other approach is to build semi-external data structures 
(see~\cite{GM13,GMCTW14} and references therein).

\section{Suffix array with deep buckets}

The mentioned idea of Manber and Myers with precomputed interval (bucket) 
boundaries for $k$ starting symbols tends to bring more gain with growing $k$, 
but also precomputing costs grow exponentially. 
Obviously, $\sigma^k$ integers are needed to be kept in the lookup table.
Our proposal is to apply hashing on relatively long strings, with an extra trick 
to reduce the number of 
unnecessary references to the text.

\begin{figure}
\begin{small}
\begin{code}
HT\_build($T[0 \ldots n-1]$, $SA[0 \ldots n-1]$, $k$, $z$, $h(.)$) \\
Precondition: $k \geq 2$ \\
\rule{\textwidth}{0.3mm} \\
\vspace{-20mm}%
(01) \startindent allocate $HT[0 \ldots z-1]$ \\
(02) \startindent \ufor $j \uassign 0$ \uto $z - 1$ \udo $HT[j] \uassign NIL$ \\
(03) \startindent $prevStr \uassign \varepsilon$ \\
(04) \startindent $j \uassign NIL$ \\
(05) \startindent $left \uassign NIL$; $right \uassign NIL$ \\
(06) \startindent \ufor $i \uassign 0$ \uto $n - 1$ \udo \\
(07) \startindent \>\>\uif $SA[i] \geq n - k$ \uthen \ \ucontinue \\
(08) \startindent \>\>\uif $T[SA[i] \ldots SA[i]+k-1] \neq prevStr$ \uthen \\
(09) \startindent \>\>\>\uif $j \neq NIL$ \uthen \\
(10) \startindent \>\>\>\>$right \uassign i-1$ \\
(11) \startindent \>\>\>\>$HT[j] \uassign (left, right)$ \\
(12) \startindent \>\>\>$j \uassign h(T[SA[i] \ldots SA[i]+k-1])$ \\
(13) \startindent \>\>\>$prevStr \uassign T[SA[i] \ldots SA[i]+k-1]$ \\
(14) \startindent \>\>\>\urepeat \\
(15) \startindent \>\>\>\>\uif $HT[j] = NIL$ \uthen \\
(16) \startindent \>\>\>\>\>$left \uassign i$ \\
(17) \startindent \>\>\>\>\>\ubreak \\
(18) \startindent \>\>\>\>\uelse $j \uassign (j + 1) \umod z$ \\
(19) \startindent \>\>\>\uuntil false \\
(20) \startindent $HT[j] \uassign (left, n-1)$ \\
(21) \startindent \ureturn $HT$ \\
\end{code}
\caption{Building the hash table of a given size $z$}
\label{fig:HT_build}
\end{small}
\end{figure}

We start with building the hash table HT (Fig.~\ref{fig:HT_build}).
The hash function is calculated for the {\em distinct} $k$-symbol ($k \geq 2$) prefixes of 
suffixes from the (previously built) suffix array.
That is, we process the suffixes in their SA order and if the current suffix 
shares its $k$-long prefix with its predecessor, it is skipped (line~08).
The value written to HT (line~11) is a pair: 
(the position in the SA of the first suffix with the 
given prefix, the position in the SA of the last suffix with the given prefix).
Linear probing is used as the collision resolution method.
As for the hash function, we used 
xxhash (\url{https://code.google.com/p/xxhash/}). 
We tested also a few alternatives: MurmurHash 
(\url{http://en.wikipedia.org/wiki/MurmurHash}) is practically as good  
as xxhash, 
CRC (\url{http://rosettacode.org/wiki/CRC-32})
is slightly slower overall (with up to about 3\% slower searches), 
while the loss of sdbm (\url{http://www.cse.yorku.ca/~oz/hash.html}) 
is greater, often exceeding 10\%.

Fig.~\ref{fig:Pattern_search} presents the pattern search (locate) procedure.
It is assumed that the pattern length $m$ is not less than $k$.
First the range of rows in the suffix array corresponding to the first two 
symbols of the pattern is found in a ``standard'' lookup table (line~1); 
an empty range immediately terminates the search with no matches returned (line~2).
Then, the hash function over the pattern prefix is calculated and a scan over the 
hash table performed until no extra collisions (line~5; return no matches) 
or found a match over the pattern prefix, which give us information about the range 
of suffixes starting with the current prefix (line~6).
In this case, the binary search strategy is applied to narrow down the SA interval 
to contain exactly the suffixes starting with the whole pattern.
(As an implementation note: the binary search could be modified to ignore the first $k$ 
symbols in the comparisons, but it did not help in our experiments, 
due to specifics of the used A\_strcmp function from the asmlib library\footnote{http://www.agner.org/optimize/asmlib.zip, v2.34, by Agner Fog.}).

\begin{figure}
\begin{small}
\begin{code}
Pattern\_search($T[0 \ldots n-1]$, $SA[0 \ldots n-1]$, $HT[0 \ldots z-1]$, $k$, $h(.)$, \\
\>\>\>$P[0 \ldots m-1]$) \\
Precondition: $m \geq k \geq 2$ \\
\rule{\textwidth}{0.3mm} \\
(1) \startindent $beg, end \uassign LUT_2[p_0,p_1]$ \\
(2) \startindent if $end < beg$ \uthen report no matches; \ureturn \\
(3) \startindent $j \uassign h(P[0 \ldots k-1])$ \\
(4) \startindent \urepeat \\
(5) \startindent \>\>\uif $HT[j] = NIL$ \uthen report no matches; \ureturn \\
(6) \startindent \>\>\uif ($beg \leq HT[j].left \leq end$) \textbf{and} \\
\>\>\>($T[SA[HT[j].left] \ldots SA[HT[j].left] + k - 1] = P[0 \ldots k-1]$) \\
(7) \startindent \>\>\>\uthen binSearch($P[0 \ldots m-1], HT[j].left, HT[j].right$); \ureturn \\
(8) \startindent \>\>$j \uassign (j + 1) \umod z$ \\
(9) \startindent \uuntil false \\
\end{code}
\caption{Pattern search with SA-hash}
\label{fig:Pattern_search}
\end{small}
\end{figure}

\subsection{Reducing the memory for the hash table}

Each slot in the hash table (HT) contains two 32-bit integers, 
for the start and the end position of the range of suffixes starting 
with the corresponding prefix of length $k$.
Yet, it is possible in practice to reduce the second value to 16 bits.
To this end, we make use of a lookup table over 
pairs of symbols (LUT2)
to initially narrow down the interval related to which the range in the HT 
will be encoded.
Then the actual range will be written approximately, with quantized
right boundary of the range.

For clarity, let us denote the new hash table with $HT_{approx}$. 
The code for building $HT_{approx}$ 
is shown in Fig.~\ref{fig:HT_approx_build}.

Let us explain why a similar saving cannot be applied also to the 
start position of the range.
This is because a collision which (unluckily) points to a subrange of 
the actual HT range that we are looking for could not be detected.
Here is an example.
Let us assume that we have two $k$-long prefixes: 
``somethin'' and ``once in'', which have the same hash value (collision).
The SA range for ``something'' is $[30200, 30700]$ and 
LUT2 stores (for ``so'') the range $[30000, 31000]$.
The SA range for ``once in'' is $[10300, 10600]$ and 
LUT2 table stores (for ``on'') the range $[10000, 11000]$.
Now we are decoding the range of ``somethin'' suffixes and 
there is a collision with ``once in''. 
Hence we obtained the SA range $[30300, 30600]$ which is a subrange of 
$[30200, 30700]$ and we cannot detect a collision. 
We are searching in narrower range, so the results may be wrong.
Quantizing only the right boundary of the range does not imply 
a similar problem.

\begin{figure}
\begin{small}
\begin{code}
HT\_approx\_build($T[0 \ldots n-1]$, $SA[0 \ldots n-1]$, $k$, $z$, $h(.)$) \\
Precondition: $k \geq 2$ \\
\rule{\textwidth}{0.3mm} \\
\vspace{-20mm}%
(01) \startindent allocate $HT_{approx}[0 \ldots z-1]$ \\
(02) \startindent \ufor $j \uassign 0$ \uto $z - 1$ \udo $HT_{approx}[j] \uassign NIL$ \\
(03) \startindent $prevStr \uassign \varepsilon$ \\
(04) \startindent $j \uassign NIL$ \\
(05) \startindent $left \uassign NIL$; $right \uassign NIL$; $beg \uassign NIL$; $end \uassign NIL$ \\
(06) \startindent \ufor $i \uassign 0$ \uto $n - 1$ \udo \\
(07) \startindent \>\>\uif $SA[i] \geq n - k$ \uthen \ \ucontinue \\
(08) \startindent \>\>\uif $T[SA[i] \ldots SA[i]+k-1] \neq prevStr$ \uthen \\
(09) \startindent \>\>\>\uif $j \neq NIL$ \uthen \\
(10) \startindent \>\>\>\>$right \uassign \lceil (i-1-beg)/step \rceil$ \\
(11) \startindent \>\>\>\>$HT_{approx}[j] \uassign (left, right)$ \\
(12) \startindent \>\>\>$j \uassign h(T[SA[i] \ldots SA[i]+k-1])$ \\
(13) \startindent \>\>\>$prevStr \uassign T[SA[i] \ldots SA[i]+k-1]$ \\
(14) \startindent \>\>\>\urepeat \\
(15) \startindent \>\>\>\>\uif $HT_{approx}[j] = NIL$ \uthen \\
(16) \startindent \>\>\>\>\>$beg, end \uassign LUT_2[p_0, p_1]$ \\
(17) \startindent \>\>\>\>\>$step \uassign \lceil (end+1-beg)/2^{16} \rceil$ \\
(18) \startindent \>\>\>\>\>$left \uassign i$ \\
(19) \startindent \>\>\>\>\>\ubreak \\
(20) \startindent \>\>\>\>\uelse $j \uassign (j + 1) \umod z$ \\
(21) \startindent \>\>\>\uuntil false \\
(22) \startindent $HT_{approx}[j] \uassign (left, \lceil (n-1-beg)/step \rceil)$ \\
(23) \startindent \ureturn $HT$ \\
\end{code}
\caption{Building the hash table with reduced memory}
\label{fig:HT_approx_build}
\end{small}
\end{figure}

\section{Fixed Block based Compact Suffix Array} \label{sec:fbsa}

M{\"a}kinen's compact suffix
array~\cite{DBLP:conf/cpm/Makinen00,DBLP:journals/fuin/Makinen03} 
finds and succinctly represents repeating suffix areas.
We propose a variant of this index
whose key feature is 
finding {\em approximate} repetitions of suffix areas of predefined size.
Choosing the fixed area size 
allows to maintain a byte-aligned data layout, 
beneficial for speed and simplicity.
Even more, by setting a natural restriction on one of the key parameters 
we force the structure's building bricks to be multiples of 32 bits, 
which prevents misaligned access to data.

M{\"a}kinen's index was the first {\em opportunistic} scheme for 
compressing a suffix array, that is such that uses less space on compressible 
texts.
The key idea was to exploit runs in the SA, that is, maximal 
segments $SA[i \ldots i + \ell - 1]$ for which there exists another 
segment $SA[j \ldots j + \ell - 1]$, such that $SA[j + s] = SA[i + s] + 1$ 
for all $0 \leq s < \ell$.
This structure still allows for binary search, only the accesses to SA cells 
require local decompression.
In our approach we take suffix areas of fixed size, e.g., 32 bytes: 
$SA[i \ldots i+31]$, and find for them other suffix array segments 
$SA[j \ldots j + \ell - 1]$ such that for each $s \in \{0, 1, \ldots, \ell - 1\}$ there exists $m \in \{0, 1, \ldots, 31\}$ such that $SA[j + s] + 1 = SA[i + m]$. 
Moreover, the sequence of chosen values of $m$ is ascending. 
%

\begin{figure}
\begin{small}
\begin{code}
FBCSA\_build($SA[0 \ldots n-1]$, $T^{BWT}$, $bs$, $ss$) \\
\rule{\textwidth}{0.3mm} \\
/* assume $n$ is a multiple of $bs$ */ \\
(01) \startindent $arr_1 \uassign [\ ]$; $arr_2 \uassign [\ ]$ \\
(02) \startindent $j \uassign 0$ \\
(03) \startindent \urepeat \\
\>\>/* current block of the suffix array is $SA[j \ldots j+bs-1]$ */ \\
(04) \startindent \>\>find 3 most frequent symbols in $T^{BWT}[j \ldots j+bs-1]$ \\
\>\>\> and store them in $MFS[0 \ldots 2]$ \\ 
\>\>\> /* if there are less than 3 distinct symbols in $T^{BWT}[j \ldots j+bs-1]$, \\ 
\>\>\>\> the trailing cells of $MFS[0 \ldots 2]$ are set to $NIL$) */ \\
(05) \startindent \>\>\ufor $i \uassign 0$ \uto $bs-1$ \udo \\
(06) \startindent \>\>\>\uif $T^{BWT}[j+i] = MFS[0]$ \uthen $arr_1$.append(00) \\
(07) \startindent \>\>\>\uelse \ \uif $T^{BWT}[j+i] = MFS[1]$ \uthen $arr_1$.append(01) \\
(08) \startindent \>\>\>\>\uelse \ \uif $T^{BWT}[j+i] = MFS[2]$ \uthen $arr_1$.append(10) \\
(09) \startindent \>\>\>\>\>\uelse $arr_1$.append(11) \\
(10) \startindent \>\>$pos_0 = T^{BWT}[j \ldots j+bs-1]$.pos($MFS[0]$) \\
(11) \startindent \>\>$pos_1 = T^{BWT}[j \ldots j+bs-1]$.pos($MFS[1]$) /* set $NIL$ if $MFS[1] = NIL$ */ \\
(12) \startindent \>\>$pos_2 = T^{BWT}[j \ldots j+bs-1]$.pos($MFS[2]$) /* set $NIL$ if $MFS[2] = NIL$ */ \\
(13) \startindent \>\>$a2s = |arr_2|$ \\
(14) \startindent \>\>$arr_2$.append($SA^{-1}[SA[j + pos_0] - 1]$) \\
(15) \startindent \>\>$arr_2$.append($SA^{-1}[SA[j + pos_1] - 1]$) /* append $-1$ if $pos_1 = NIL$ */ \\
(16) \startindent \>\>$arr_2$.append($SA^{-1}[SA[j + pos_2] - 1]$) /* append $-1$ if $pos_2 = NIL$ */ \\
(17) \startindent \>\>\ufor $i \uassign 0$ \uto $bs-1$ \udo \\
(18) \startindent \>\>\>\uif ($T^{BWT}[j+i] \not\in \{MFS[0], MFS[1], MFS[2]\}$) \textbf{or} ($SA[j+i] \umod ss = 0$) \\ 
(19) \startindent \>\>\>\>\uthen $arr_1$.append(1); $arr_2$.append($SA[j+i]$) \\
(20) \startindent \>\>\>\uelse $arr_1$.append(0) \\
(21) \startindent \>\>$arr_1$.append($a2s$) \\
(22) \startindent \>\>$j \uassign j + bs$ \\
(23) \startindent \>\>\uif $j = n$ \uthen \ \ubreak \\
(24) \startindent \uuntil false \\
(25) \startindent \ureturn $(arr_1, arr_2)$ \\
\end{code}
\caption{Building the fixed block based compact suffix array (FBCSA)}
\label{fig:FBSA_build}
\end{small}
\end{figure}

The construction algorithm for our structure, called {\em fixed block based 
compact suffix array} (FBCSA), is presented in Fig.~\ref{fig:FBSA_build}.
As a result, we obtain two arrays, $arr_1$ and $arr_2$, which are empty 
at the beginning, and their elements are always appended at the end 
during the construction.
The elements appended to $arr_1$ are single bits or pairs of bits 
while $arr_2$ stores suffix array indexes (32-bit integers).

The construction makes use of the suffix array $SA$ of text $T$,
the inverse suffix array $SA^{-1}$ and $T^{BWT}$ (which can be obtained 
from $T$ and $SA$, that is, $T^{BWT}[i] = T[(SA[i] - 1) \mmod n]$).

Additionally, there are two construction-time parameters: 
block size $bs$ and sampling step $ss$.
The block size tells how many successive $SA$ indexes are encoded together 
and is assumed to be a multiple of 32, for int32 alignment of the structure 
layout.
The parameter $ss$ means that every $ss$-th $SA$ index will be 
represented verbatim.
This sampling parameter is a time-space tradeoff; 
using larger $ss$ reduces the overall space but decoding a particular SA index 
typically involves more recursive invocations.

Let us describe the encoding procedure for one block,
$SA[j \ldots j+bs-1]$, where $j$ is a multiple of $bs$.

First we find the three most frequent symbols in $T^{BWT}[j \ldots j+bs-1]$ and store them 
(in arbitrary order) in a small helper array $MFS[0 \ldots 2]$ (line~04).
If the current block of $T^{BWT}$ does not contain three different symbols, 
the $NIL$ value will be written in the last one or two cell(s) of $MFS$.
Then we write information about the symbols from $MFS$ in the current block of $T^{BWT}$
into $arr_1$: we append 2-bit combination (00, 01 or 10) if a given symbol is from $MFS$ 
and the remaining combination (11) otherwise (lines~05--09).
We also store the positions of the first occurrences of the symbols from $MFS$ 
in the current block of $T^{BWT}$, using the variables $pos_0$, $pos_1$, $pos_2$ 
(lines~10--12); 
again $NIL$ values are used if needed.
These positions allow to use links to runs of suffixes 
preceding subsets of the current ones marked by the respective symbols from $MFS$.

We believe that a small example will be useful here.
Let $bs = 8$ and the current block be $SA[400 \ldots 407]$ (note this is a toy example 
and in the real implementation $bs$ must be a multiple of 32).
The $SA$ block contains the indexes: 1000, 522, 801, 303, 906, 477, 52, 610.
Let their preceding symbols (from $T^{BWT}$) be: $a$, $b$, $a$, $c$, $d$, $d$, $b$, $b$.
The three most frequent symbols, written to $MFS$, are thus: $b$, $a$, $d$.
The first occurrences of these symbols are at positions: 
$pos_0 = 1$, $pos_1 = 0$ and $pos_2 = 4$. 
We conclude that SA has the following groups of suffix offsets: 
$[521, 51, 609]$ (as there are three symbols $b$ in the current block of $T^{BWT}$), 
$[999, 800]$ and $[905, 476]$ and they start at positions: 
$SA^{-1}[521]$, $SA^{-1}[999]$ and $SA^{-1}[905]$.

We come back to the pseudocode.
The described (up to three) links are obtained thanks to 
$SA^{-1}$ (lines~14--16) and are written to $arr_2$.
Finally, the offsets of the suffixes preceded with a symbol not from $MFS$ 
(if any) have to be written to $arr_2$ explicitly.
Additionally, the sampled suffixes (i.e., those whose offset modulo $ss$ is 0) 
are handled in the same way (line~18).
To distinguish between referrentially encoded and explicitly written 
suffix offsets, we spent a bit per suffix and append them to $arr_1$ 
(lines~19--20).
To allow for easy synchronization between the portions of data in $arr_1$ 
and $arr_2$, the size of $arr_2$ (in bytes) as it was before processing the 
current block is written to $arr_1$ (line~21).

\begin{figure}
\begin{code}
Find($arr_1$, $arr_2$, $bs$, $i$) \\
\rule{\textwidth}{0.3mm} \\
/* assume $bs$ is a multiple of 32 */ \\
(01) \startindent $of_1 \uassign bs / 16$ \\
(02) \startindent $of_2 \uassign (bs / 16) + (bs / 32)$ \\
(03) \startindent $cb_{start} \uassign \floor{i / bs} * (of_2 + 1)$ \\
(04) \startindent $cb_{currpos} \uassign i\ \%\ bs$ \\
(05) \startindent $d_0 \uassign \floor{cb_{currpos} / 32}$ \\
(06) \startindent $c \uassign arr_1[cb_{start} + of_{2}]$ \\
(07) \startindent $bBits \uassign int2bits(arr_1[cb_{start} + of_{1} \ldots cb_{start} + of_{1} + d_0])$ \\
(08) \startindent \uif $bBits[cb_{currpos}] = 1$ \uthen \\
(09) \startindent \>\>\ureturn $arr_2[c + 3 + popc_1(bBits[0 \dots cb_{currpos} - 1])]$ \\
(10) \startindent \uelse \\
(11) \startindent \>\>$d_1 \uassign \floor{cb_{currpos} / 16}$ \\
(12) \startindent \>\>$dBits \uassign int2dibits(arr_1[cb_{start} \ldots cb_{start} + d_1])$ \\
(13) \startindent \>\>$sym \uassign dBits[cb_{currpos}]$ \\
(14) \startindent \>\>\ureturn Find($arr_1$, $arr_2$, $bs$, \\
  \hspace{5.5em} $arr_2[c + int(sym)] + popc_{sym}(dBits[0 \ldots cb_{currpos}-1])) + 1$ \\
\end{code}
\caption{Find($i$) extracts $SA[i]$ from the FBCSA structure}
\label{fig:FBSA_access}
\end{figure}

Fig.~\ref{fig:FBSA_access} presents the function Find($i$), 
which returns $SA[i]$.
The helper arrays $bBits$ and $dBits$ contain respectively bits and 
pairs of bits (extracted from one or several integers).
The function $popc_c$ (popcount) returns the number of occurrences 
of symbol (integer) $c$ in the given array of symbols (integers).
In modern CPUs $popc_1$ for a bit-vector of size e.g. 64 is usually 
available as a single op-code.

\section{Experimental results}

All experiments were run on a machine equipped with a 6-core Intel i7 CPU
(4930K) clocked at 3.4\,GHz, with 64\,GB of RAM, 
running Ubuntu 14.04 LTS 64-bit.
The RAM modules were $8 \times 8$\,GB DDR3-1600 with the timings 11-11-11 
(Kingston KVR16R11D4K4/64).
The CPU cache sizes were:
$6 \times 32$\,KB (data) and $6 \times 32$\,KB (instructions) in the L1 level,
$6 \times 256$\,KB in L2 and 12\,MB in L3.
One CPU core was used for the computations.
All codes were written in C++ and compiled with 64-bit gcc 4.8.2, 
with \texttt{-O3} option
(and for the FBCSA search algorithms with the additional \texttt{-mpopcnt} option).
The source codes for the FBCSA algorithm can be downloaded from \url{http://ranisz.iis.p.lodz.pl/indexes/fbcsa/}.

The test datasets were taken from the popular Pizza~\&~Chili 
site (\url{http://pizzachili.dcc.uchile.cl/}).
For most experiments we used the 200-megabyte versions of the files \texttt{dna},
\texttt{english}, \texttt{proteins}, \texttt{sources} and \texttt{xml}.
Only to compare search times of FBCSA variants against M{\"a}kinen's CSA
we used 50-megabyte datasets, due to text size limitations of the 
CSA implementation.

In order to test the search algorithms, we generated 500 thousand patterns 
for each used pattern length; the patterns were extracted randomly from the 
corresponding datasets (i.e., each pattern returns at least one match).

In the first experiment we compared pattern search (count) speed 
using the following indexes:
\begin{itemize}
\item plain suffix array (SA),
\item suffix array with a lookup table over the first 2 symbols (SA-LUT2),
\item suffix array with a lookup table over the first 3 symbols (SA-LUT3),
\item the proposed suffix array with deep buckets, 
      with hashing the prefixes of length $k = 8$ 
      (only for \texttt{dna} $k = 12$ and for \texttt{proteins} $k = 5$ is used); 
      the load factor $\alpha$ in the hash table was set to 90\% (SA-hash),
\item a more compact variant of SA-hash, with 6 bytes rather than 8 bytes per 
      entry in the hash table (SA-hash-dense),
\item the proposed fixed block based compact suffix array with parameters $bs = 32$ 
      and $ss = 5$ (FBCSA),
\item FBCSA (parameters as before) with a lookup table over the first 2 symbols 
      (FBCSA-LUT2),
\item FBCSA (parameters as before) with a lookup table over the first 3 symbols 
      (FBCSA-LUT3),
\item FBCSA (parameters as before) with a hash of prefixes of length $k = 8$ 
      (only for \texttt{dna} $k = 12$ and for \texttt{proteins} $k = 5$ is used);
      the load factor in the hash table was set to 90\% (FBCSA-hash),
\item a more compact variant of FBCSA-hash, with 6 bytes rather than 8 bytes per 
      entry in the hash table (FBCSA-hash-dense).
\end{itemize}

\begin{figure}
\centerline{
\includegraphics[width=0.49\textwidth,scale=1.0]{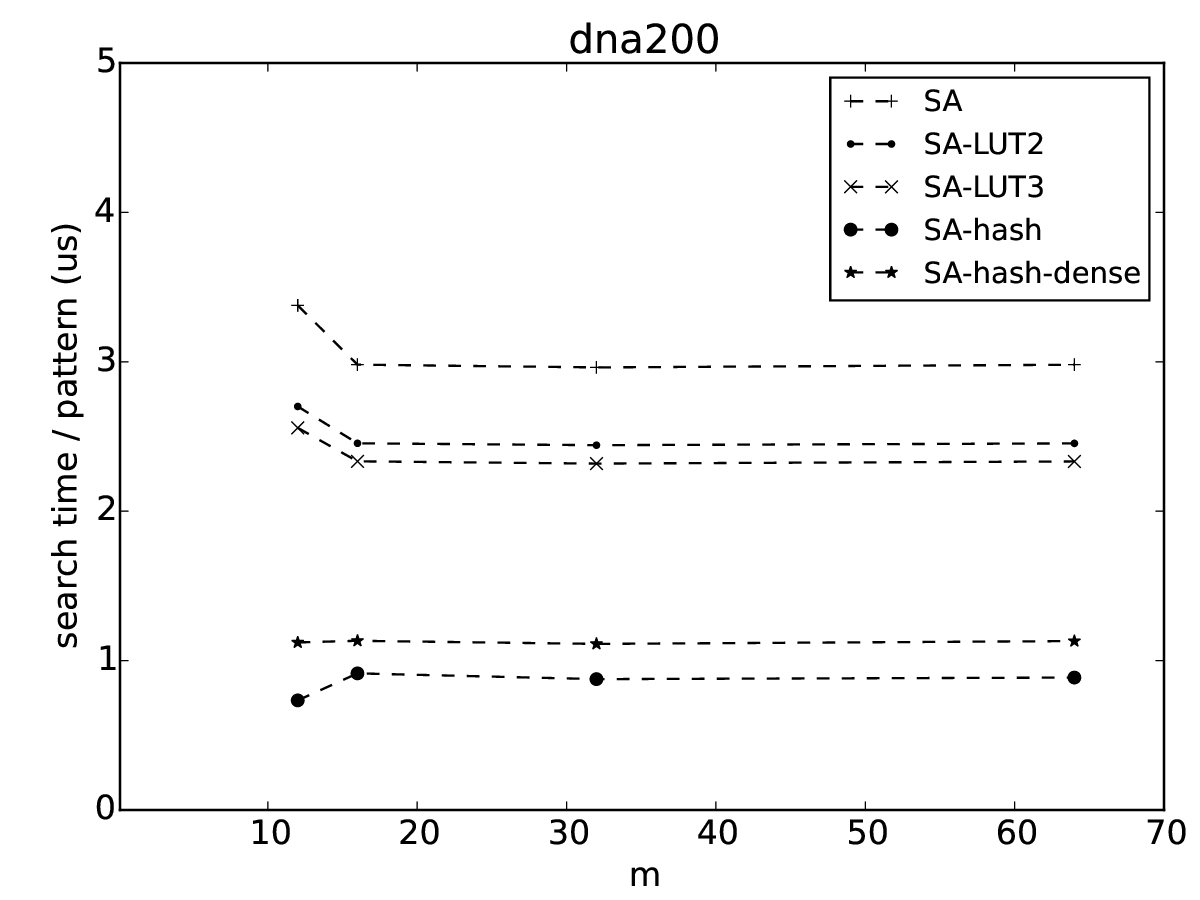}
\includegraphics[width=0.49\textwidth,scale=1.0]{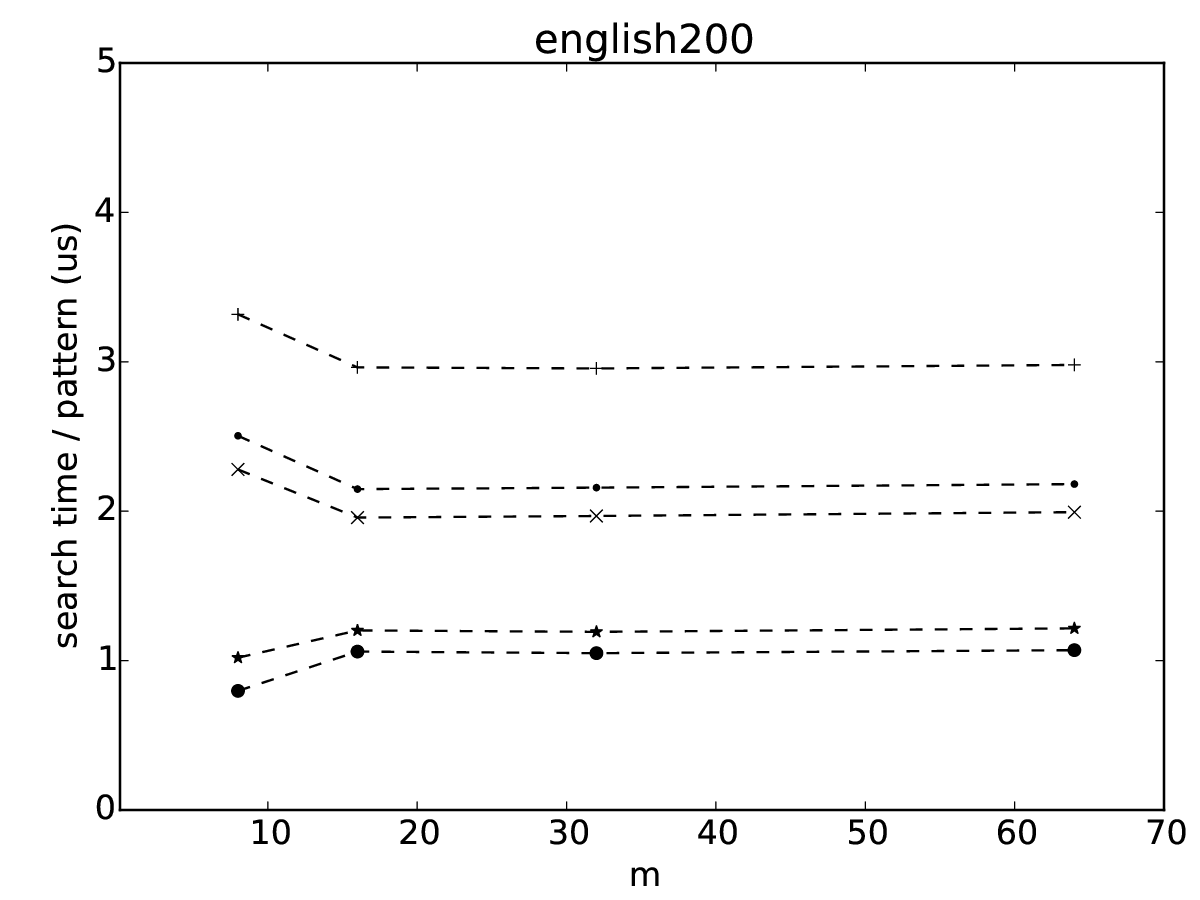}
}
\centerline{
\includegraphics[width=0.49\textwidth,scale=1.0]{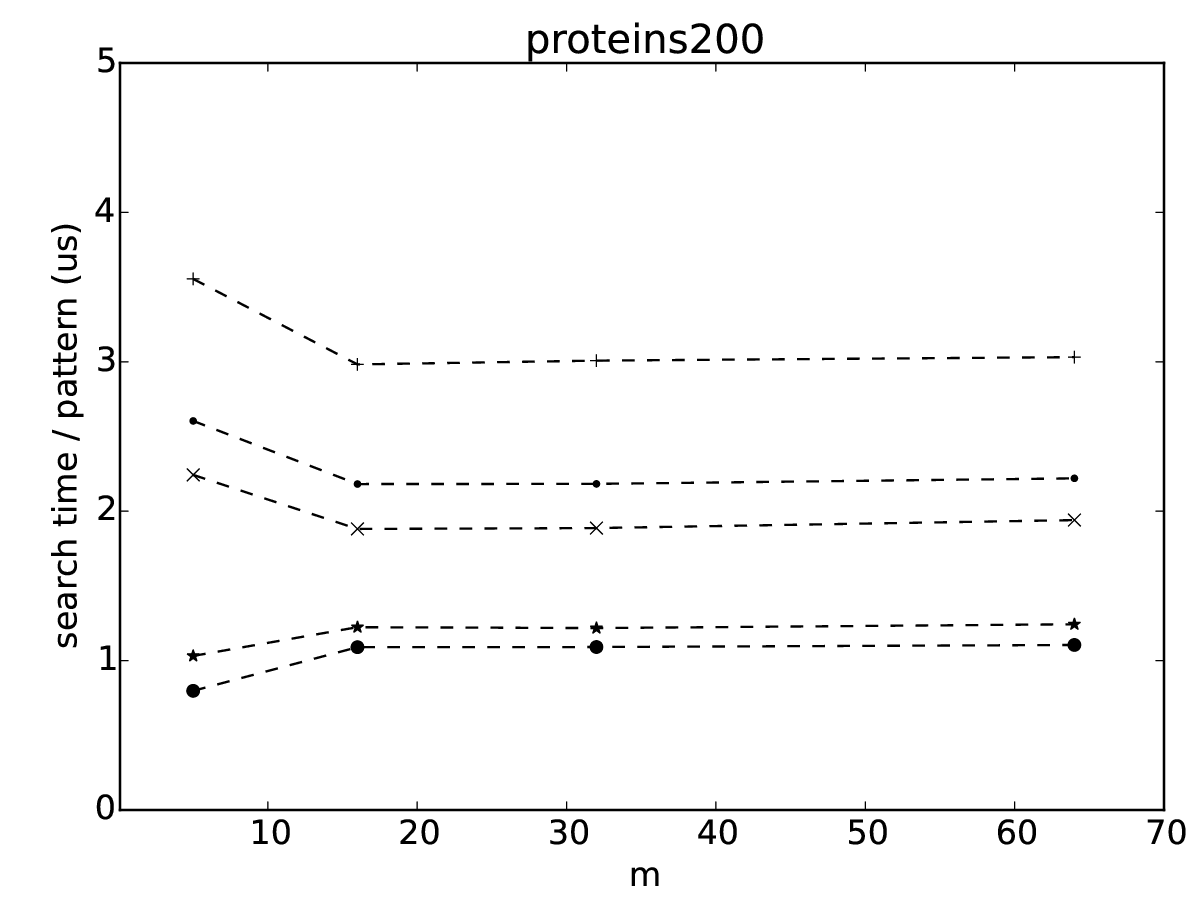}
\includegraphics[width=0.49\textwidth,scale=1.0]{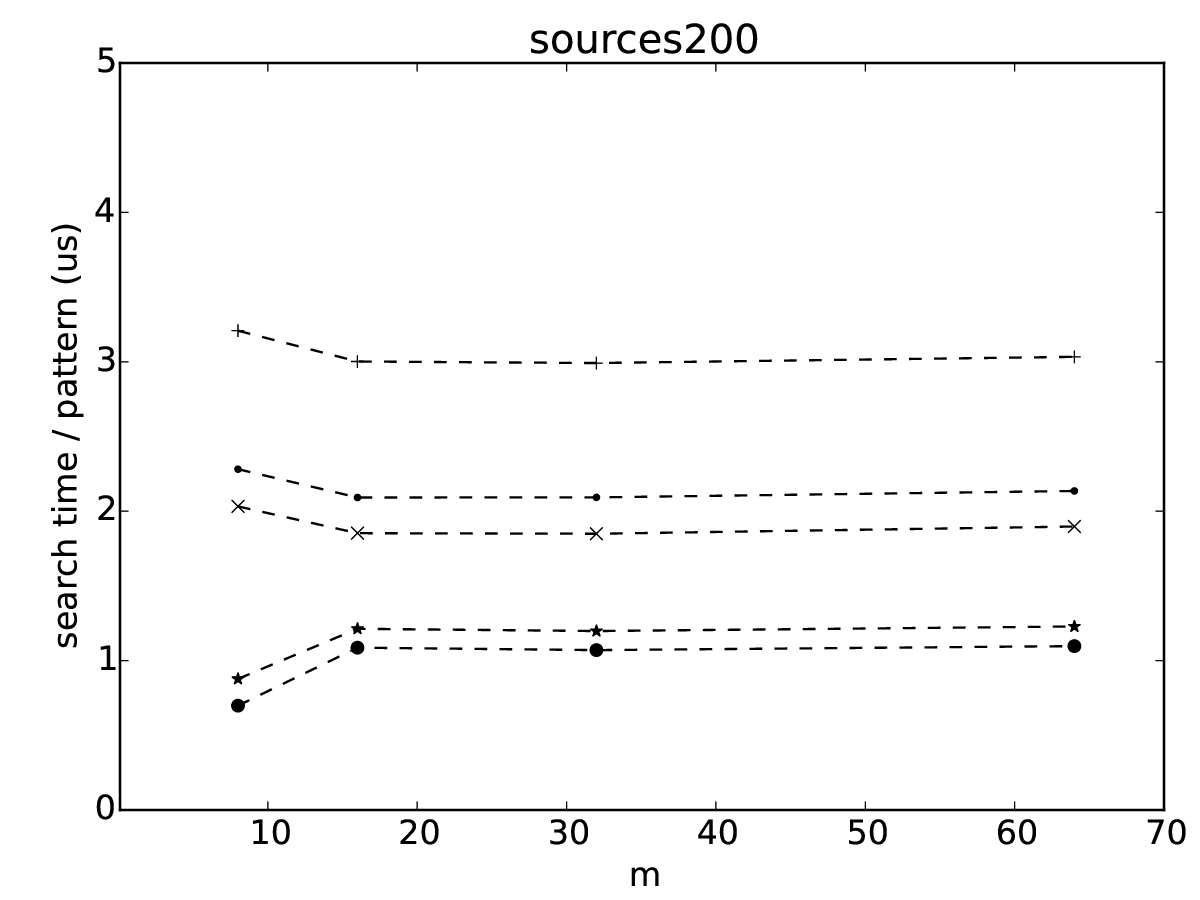}
}
\centerline{
\includegraphics[width=0.49\textwidth,scale=1.0]{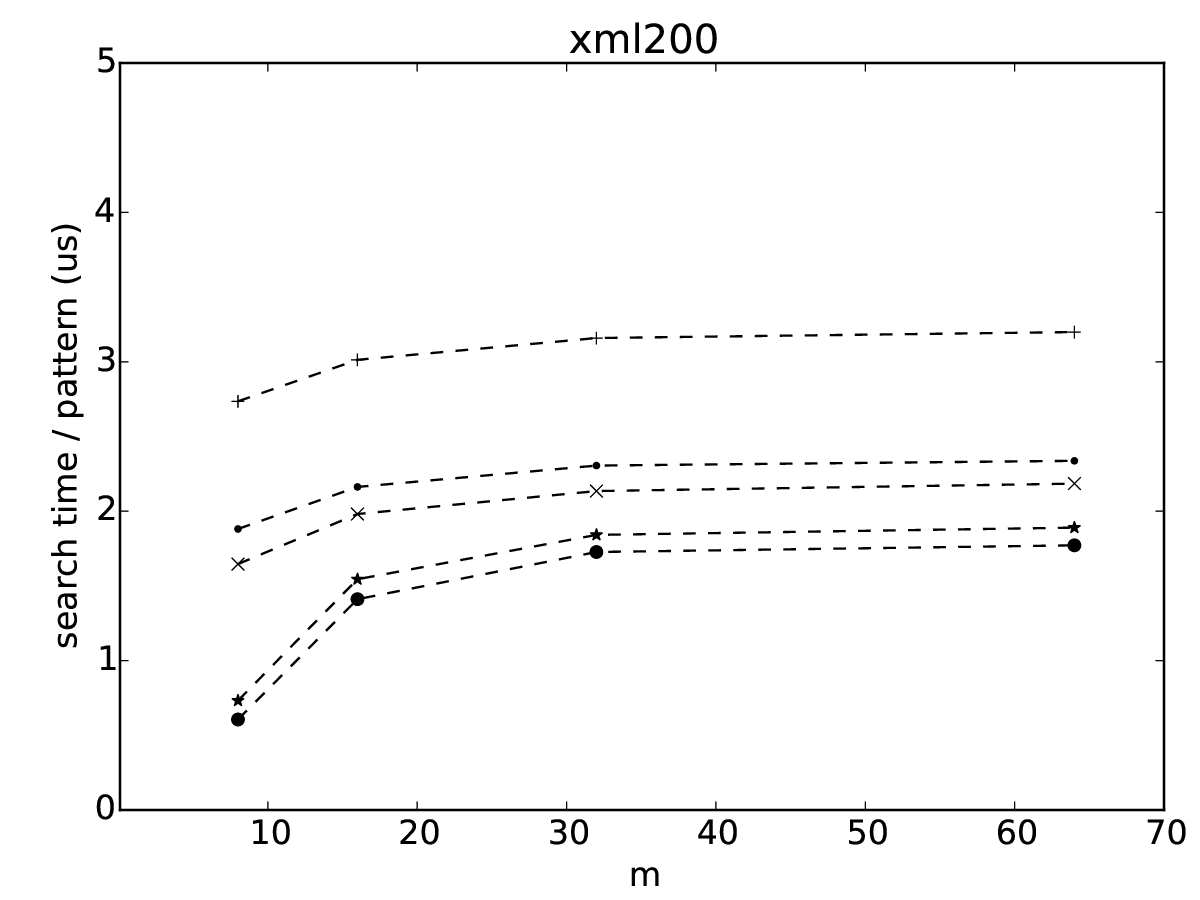}
}
\caption[Results]
{Pattern search time (count query). 
All times are averages over 500K random patterns of the same length 
$m = \{m_{min}, 16, 32, 64\}$, where $m_{min}$ is 8 for most datasets 
except for \texttt{dna} (12) and \texttt{proteins} (5). 
The patterns were extracted from the respective texts.}
\label{fig:times1a}
\end{figure}

\begin{figure}
\centerline{
\includegraphics[width=0.49\textwidth,scale=1.0]{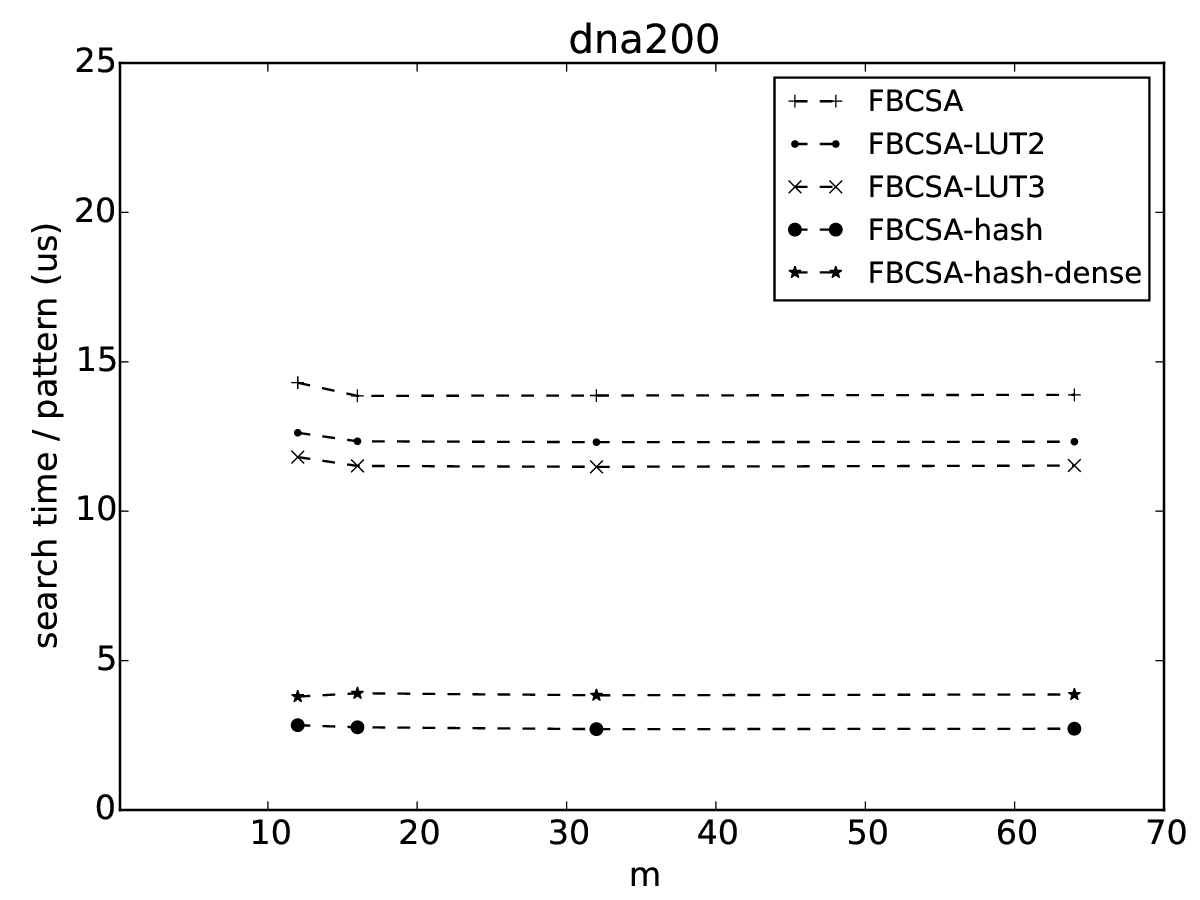}
\includegraphics[width=0.49\textwidth,scale=1.0]{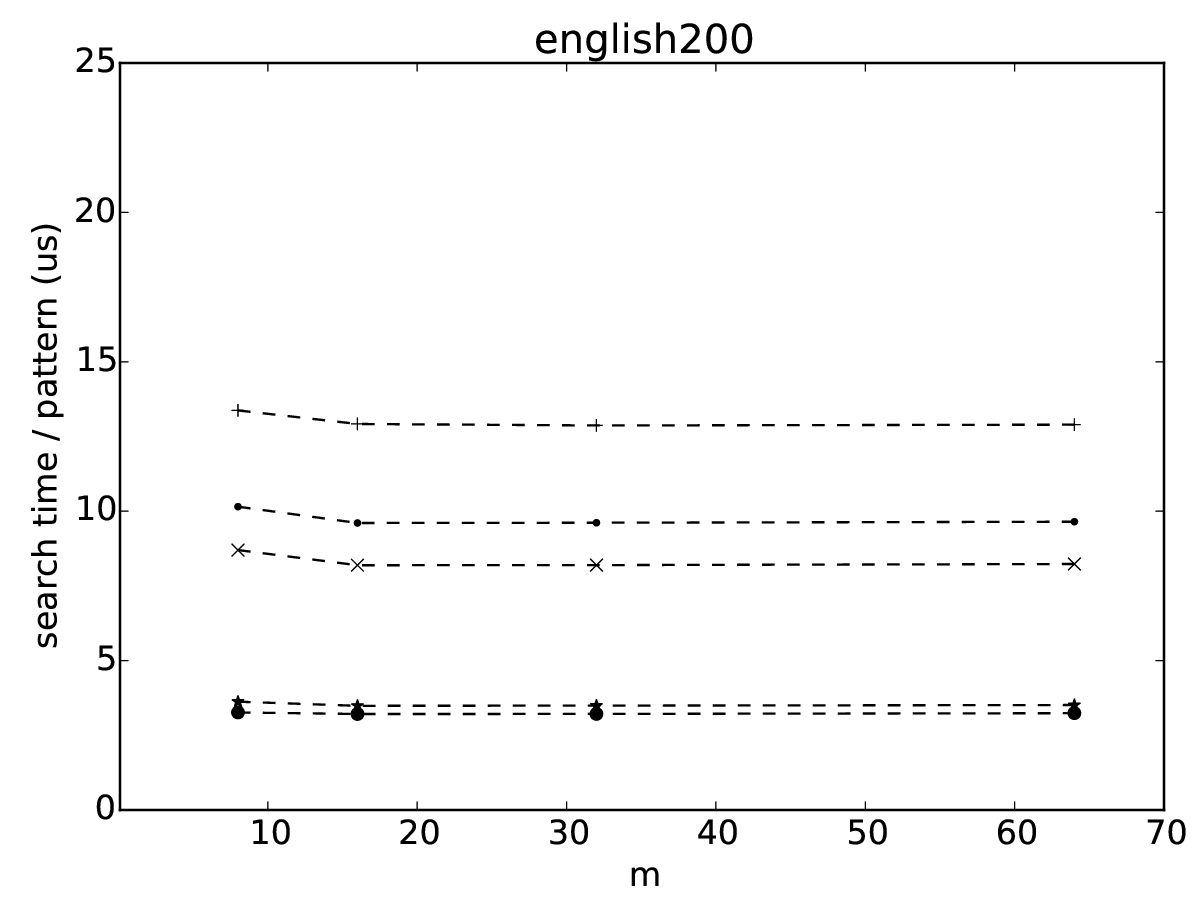}
}
\centerline{
\includegraphics[width=0.49\textwidth,scale=1.0]{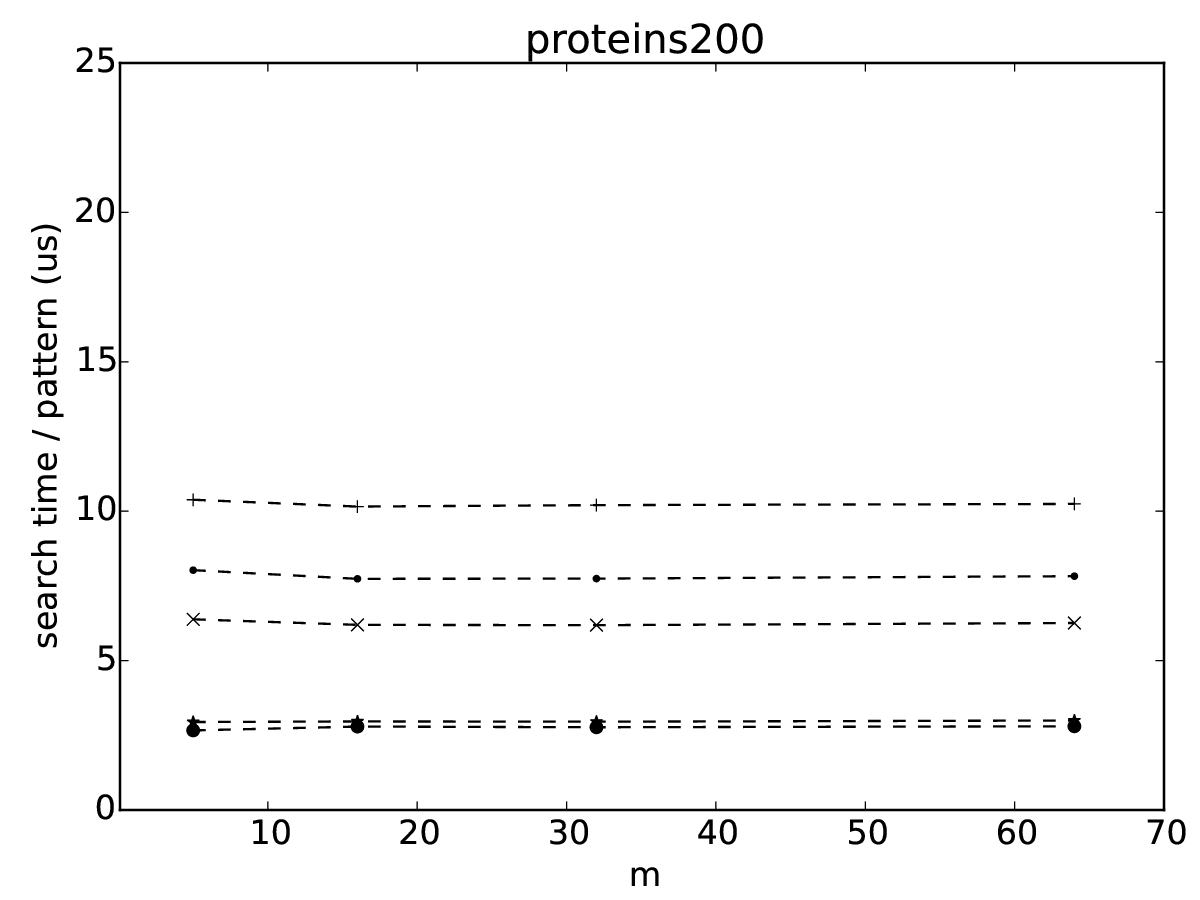}
\includegraphics[width=0.49\textwidth,scale=1.0]{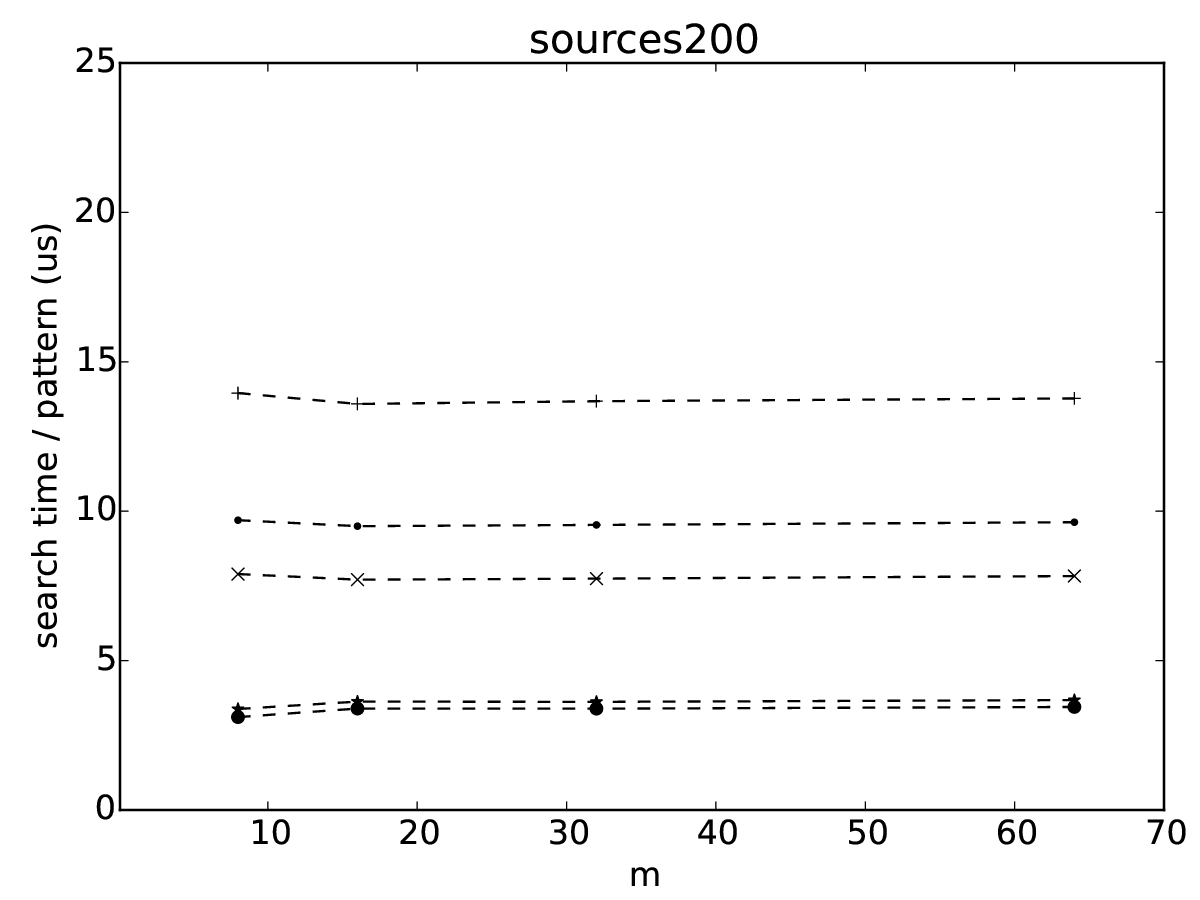}
}
\centerline{
\includegraphics[width=0.49\textwidth,scale=1.0]{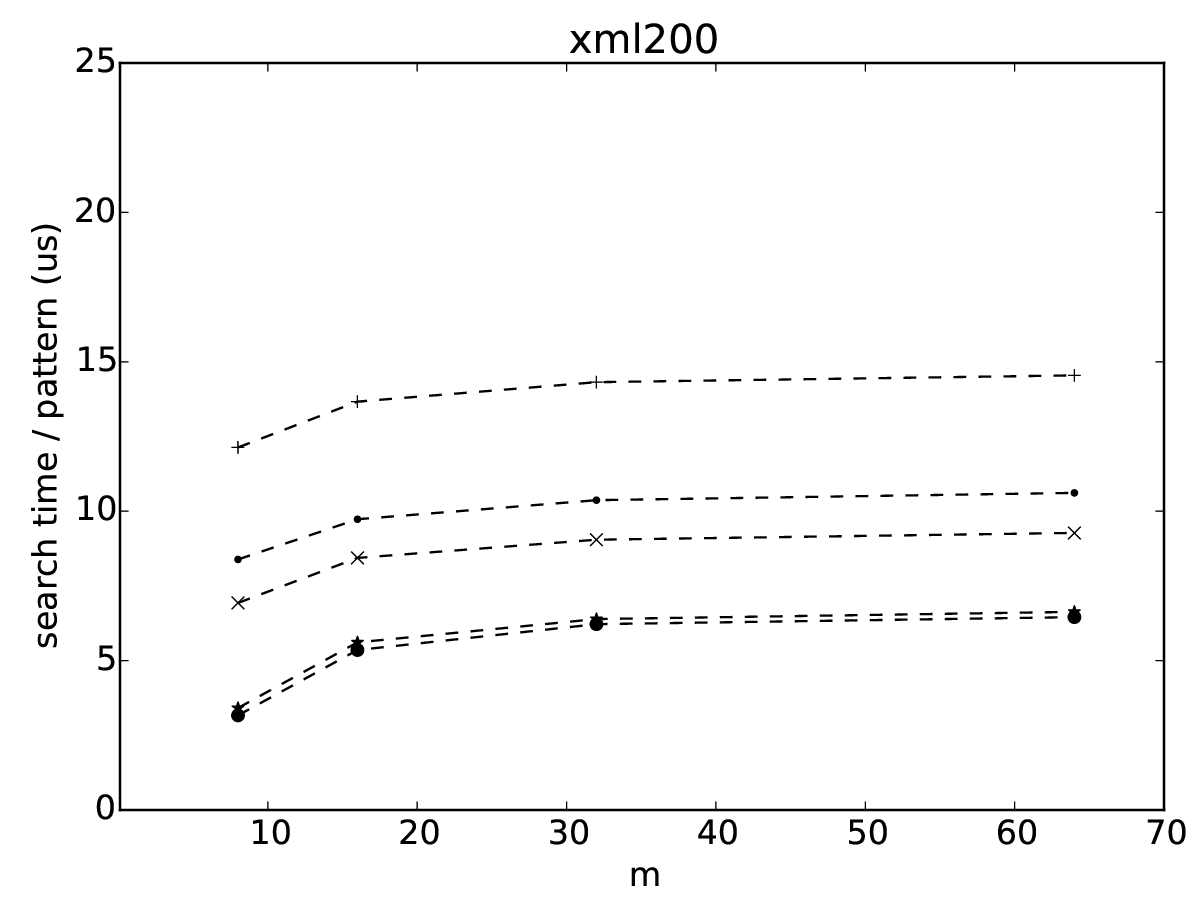}
}
\caption[Results]
{Pattern search time (count query). 
All times are averages over 500K random patterns of the same length 
$m = \{m_{min}, 16, 32, 64\}$, where $m_{min}$ is 8 for most datasets 
except for \texttt{dna} (12) and \texttt{proteins} (5). 
The patterns were extracted from the respective texts.}
\label{fig:times1b}
\end{figure}

The results are presented in Fig.~\ref{fig:times1a} (faster indexes) 
and Fig.~\ref{fig:times1b} (FBCSA variants).
As expected, SA-hash is the fastest index among the tested ones.
The reader may also look at Table~\ref{table:speedups} with a rundown 
of the achieved speedups, where the plain suffix array is 
the baseline index and its speed is denoted with 1.00.

\begin{table}
\centering
\begin{tabular}{lrrrrr}
\hline
    &~~~~~~~~\texttt{dna}~&~~\texttt{english}~&~\texttt{proteins}~&~~\texttt{sources}~&~~~~~~\texttt{xml}~\\
\hline
$m = 16$ & & & & & \\
\hline
SA    & 1.00~~& 1.00~~& 1.00~~& 1.00~~& 1.00~~\\
SA-LUT2 & 1.21~~& 1.38~~& 1.37~~& 1.44~~& 1.39~~\\
SA-LUT3 & 1.28~~& 1.51~~& 1.59~~& 1.62~~& 1.52~~\\
SA-hash & 3.26~~& 2.79~~& 2.74~~& 2.76~~& 2.14~~\\
SA-hash-dense & 2.63~~& 2.46~~& 2.44~~& 2.47~~& 1.95~~\\
\hline
$m = 64$ & & & & & \\
\hline
SA    & 1.00~~& 1.00~~& 1.00~~& 1.00~~& 1.00~~\\
SA-LUT2 & 1.21~~& 1.37~~& 1.37~~& 1.42~~& 1.37~~\\
SA-LUT3 & 1.28~~& 1.49~~& 1.56~~& 1.60~~& 1.46~~\\
SA-hash & 3.36~~& 2.78~~& 2.74~~& 2.77~~& 1.81~~\\
SA-hash-dense & 2.64~~& 2.45~~& 2.44~~& 2.47~~& 1.69~~\\

\hline
\end{tabular}
\vspace{4mm}
\caption{Speedups with regard to the search speed of the plain suffix array, 
for the five 200\,MB datasets and pattern lengths $m = 16$ and $m = 64$}
\label{table:speedups}
\end{table}

The SA-hash index has two drawbacks: it requires significantly more space 
than the standard SA and we assume (at construction time) a minimal 
pattern length $m_{min}$.
The latter issue may be eliminated, but for the price of even more space use; 
namely, we can build one hash table for each pattern length from 1 to $m_{min}$ 
(counting queries for those short patterns do not ever need to perform 
binary search over the suffix array).
For the shortest lengths ($\{1, 2\}$ or $\{1, 2, 3\}$) lookup tables may be 
alternatively used.

We have not implemented this ``all-HT'' variant, but it is easy to estimate 
the memory use for each dataset.
To this end, one needs to know the number of distinct $q$-grams for 
$q \leq m_{min}$ (Table~\ref{table:qgrams}).
Note that the alphabet size, i.e., the number of 1-grams,
for the DNA and proteins datasets is 16 and 25, respectively.
These surprisingly large values are explained by the content of the files 
in the corpus, ``polluted'' slightly with textual headers, End-of-Line symbols, etc.

\begin{table}
\centering
\begin{tabular}{lrrrrr}
\hline
~~$q$   &~~~~~~~\texttt{dna}~~&~~~~\texttt{english}~~&~~\texttt{proteins}~~&~~~\texttt{sources}~~&~~~~~~~~~~\texttt{xml}~~\\
\hline
~~1 & 16~~~& 225~~~& 25~~~& 230~~~& 96~~~\\
~~2 & 152~~~& 10,829~~~& 607~~~& 9,525~~~& 7,054~~~\\
~~3 & 683~~~& 102,666~~~& 11,607~~~& 253,831~~~& 141,783~~~\\
~~4 & 2,222~~~& 589,230~~~& 224,132~~~& 1,719,387~~~& 908,131~~~\\
~~5 & 5,892~~~& 2,150,525~~~& 3,623,281~~~& 5,252,826~~~& 2,716,438~~~\\
~~6 & 12,804~~~& 5,566,993~~~& 36,525,895~~~& 10,669,627~~~& 5,555,190~~~\\
~~7 & 28,473~~~& 11,599,445~~~& 94,488,651~~~& 17,826,241~~~& 8,957,209~~~\\
~~8 & 80,397~~~& 20,782,043~~~& 112,880,347~~~& 26,325,724~~~& 12,534,152~~~\\
~~9 & 279,680~~~& 33,143,032~~~& 117,199,335~~~& 35,666,486~~~& 16,212,609~~~\\
~10 & 1,065,613~~~& 48,061,001~~~& 119,518,691~~~& 45,354,280~~~& 20,018,262~~~\\
\hline
\end{tabular}
\vspace{4mm}
\caption{The number of distinct $q$-grams ($1 \ldots 10$) in the 200\,MB datasets.
The number of distinct 12-grams for \texttt{dna} is 13,752,341.}
\label{table:qgrams}
\end{table}

An obvious space-time factor in a hash table with open addressing 
is its load factor $\alpha$. 
We checked several values of $\alpha$ on two datasets 
(Table~\ref{table:HT_lf}) 
to conclude that using $\alpha = 90\%$ 
is a reasonable alternative 
to $\alpha = 50\%$, as the pattern search times grow by only about 10\% or less.

The number of bytes for one hash table with $z$ entries 
and $0 < \alpha \leq 1$ load factor is, 
in our implementation of SA-hash, $z \times 8 \times (1/\alpha)$, 
since each entry contains two 4-byte integers.
For example, in our experiments the hash table for \texttt{english} 
with $\alpha = 90\%$
needed 20,782,043 $\times (8/0.9) = $ 184,729,272 bytes, i.e.,  88.1\%
of the size of the text itself.
Note that the overhead in the SA-hash-dense variant with the same $\alpha$ 
is 20,782,043 $\times (6/0.9) = $ 138,546,954 bytes, i.e.,  66.1\% 
of the text size.

\begin{table}
\centering
\begin{tabular}{lccccccc}
\hline
                    & \multicolumn{7}{c}{HT load factor (\%)}  \\
\cline{2-8}
     &~~25~~&~~50~~&~~60~~&~~70~~&~~80~~&~~90~~&~~95~~ \\
\hline
\texttt{dna}, $12~~~$ & 0.625 & 0.636 & 0.643 & 0.655 & 0.672 & 0.717 & 0.789 \\
\texttt{dna}, $16~~~$ & 0.807 & 0.817 & 0.825 & 0.836 & 0.855 & 0.903 & 0.978 \\
\texttt{dna}, $32~~~$ & 0.781 & 0.792 & 0.806 & 0.809 & 0.829 & 0.877 & 0.953 \\
\texttt{dna}, $64~~~$ & 0.791 & 0.802 & 0.814 & 0.819 & 0.837 & 0.883 & 0.966 \\
\texttt{english}, $8~~~$ & 0.734 & 0.740 & 0.744 & 0.749 & 0.754 & 0.769 & 0.782 \\
\texttt{english}, $16~~~$ & 1.024 & 1.026 & 1.034 & 1.034 & 1.039 & 1.053 & 1.064 \\
\texttt{english}, $32~~~$ & 1.013 & 1.019 & 1.023 & 1.027 & 1.036 & 1.042 & 1.057 \\
\texttt{english}, $64~~~$ & 1.035 & 1.040 & 1.043 & 1.047 & 1.053 & 1.063 & 1.079 \\
\hline
\end{tabular}
\vspace{4mm}
\caption{Average pattern search times (in $\mu$s) in function of the HT load factor 
$\alpha$ for the SA-hash algorithm (xxhash function used). Each 200-megabyte  dataset name followed with the pattern length ($m$).}
\label{table:HT_lf}
\end{table}

Finally, in Table~\ref{table:space} we present the overall space use 
for the five non-compact SA variants: 
plain SA, SA-LUT2, SA-LUT3, SA-hash and SA-hash-dense, plus SA-allHT(-dense), 
which is a (not implemented) structure comprising a suffix array, a LUT2 and 
one hash table for each $k \in \{3, 4, \ldots , m_{min}\}$.
The space is expressed as a multiple of the text length $n$ (including the text), 
which is for example 5.000 for the plain suffix array.
We note that the lookup table structures become a relatively smaller fraction 
when larger texts are indexed.
For the variants with hash tables we take two load factors: 50\% and 90\%.

\begin{table}
\centering
\begin{tabular}{lrrrrr}
\hline
   &~\texttt{dna}~&~\texttt{english}~&\texttt{proteins}~&~\texttt{sources}~&~~~~~\texttt{xml}~~\\
\hline
SA    & 5.000~~& 5.000~~& 5.000~~& 5.000~~& 5.000~~\\
SA-LUT2 & 5.001~~& 5.001~~& 5.001~~& 5.001~~& 5.001~~\\
SA-LUT3 & 5.320~~& 5.320~~& 5.320~~& 5.320~~& 5.320~~\\
SA-hash-50 & 6.050~~& 6.587~~& 5.278~~& 7.010~~& 5.958~~\\
SA-hash-90 & 5.583~~& 5.882~~& 5.154~~& 6.117~~& 5.532~~\\
SA-hash-dense-90 & 5.438~~& 5.661~~& 5.116~~& 5.838~~& 5.399~~\\
SA-allHT-50 & 6.472~~& 8.114~~& 5.296~~& 9.736~~& 7.353~~\\
SA-allHT-90 & 5.818~~& 6.730~~& 5.164~~& 7.631~~& 6.307~~\\
SA-allHT-dense-90 & 5.613~~& 6.298~~& 5.123~~& 6.973~~& 5.980~~\\
\hline
\end{tabular}
\vspace{4mm}
\caption{Space use for the non-compact data structures as a multiple 
of the indexed text size (including the text), with the assumption 
that text symbols are represented in 1 byte each and SA offsets are 
represented in 4 bytes. The datasets have 200\,MB in size. 
The value of $m_{min}$ 
for SA-hash-50 and SA-hash-90, used in the construction of these structures 
and affecting their size, is like in the experiments from Fig.~\ref{fig:times1a}.
The index SA-allHT-* contains LUT2 and one hash table for each 
$k \in \{3, 4, \ldots, m_{min}\}$, when $m_{min}$ depends on the current dataset, 
as explained.
The -50 and -90 suffixes in the structure names denote the hash load factors 
(in percent).
}
\label{table:space}
\end{table}

In the next set of experiments we evaluated the FBCSA index. 
Its properties of interest, for various block size ($bs$) and 
sampling step ($ss$) parameters, are: the space use, 
pattern search times, times to access (extract) one random SA cell, 
times to access (extract) multiple consecutive SA cells. 
For $bs$ we set the values 32 and 64.
The $ss$ was tested in a wider range ($\{3, 4, 5, 8, 12, 16, 32\}$).
Using $bs = 64$ results in better compression but decoding a cell 
is also slightly slower (see Fig.~\ref{fig:times2}).

We tried to compare FBCSA against its competitors. 
Unfortunately, we were unable to run LCSA~/~LCSA-Psi~\cite{GNFjea14}
(in spite of contacting its authors) 
and MakCSA~\cite{DBLP:journals/fuin/Makinen03} cannot (directly) 
access single SA cells.
From the comparison with the results presented in~\cite[Sect.~4]{GNFjea14} 
we conclude that FBCSA is a few times faster in single cell access 
than the other related algorithms, MakCSA~\cite{DBLP:journals/fuin/Makinen03} 
(augmented with a compressed bitmap from~\cite{RamanRR02} to extract arbitrary 
ranges of the suffix array) and LCSA~/~LCSA-Psi~\cite{GNFjea14}, 
at similar or better compression.
Extracting $c$ consecutive cells is not however an efficient operation for FBCSA 
(as opposed to MakCSA and LCSA~/~LCSA-Psi, see Figs~5--7 in ~\cite{GNFjea14}), 
yet for small $ss$ the time growth is slower than linear, due to a few sampled 
(and thus written explicitly) SA offsets in a typical block (Fig.~\ref{fig:times3}).
Therefore, in extracting only 5 or 10 successive cells our index is still 
competitive.

\begin{figure}
\centerline{
\includegraphics[width=0.49\textwidth,scale=1.0]{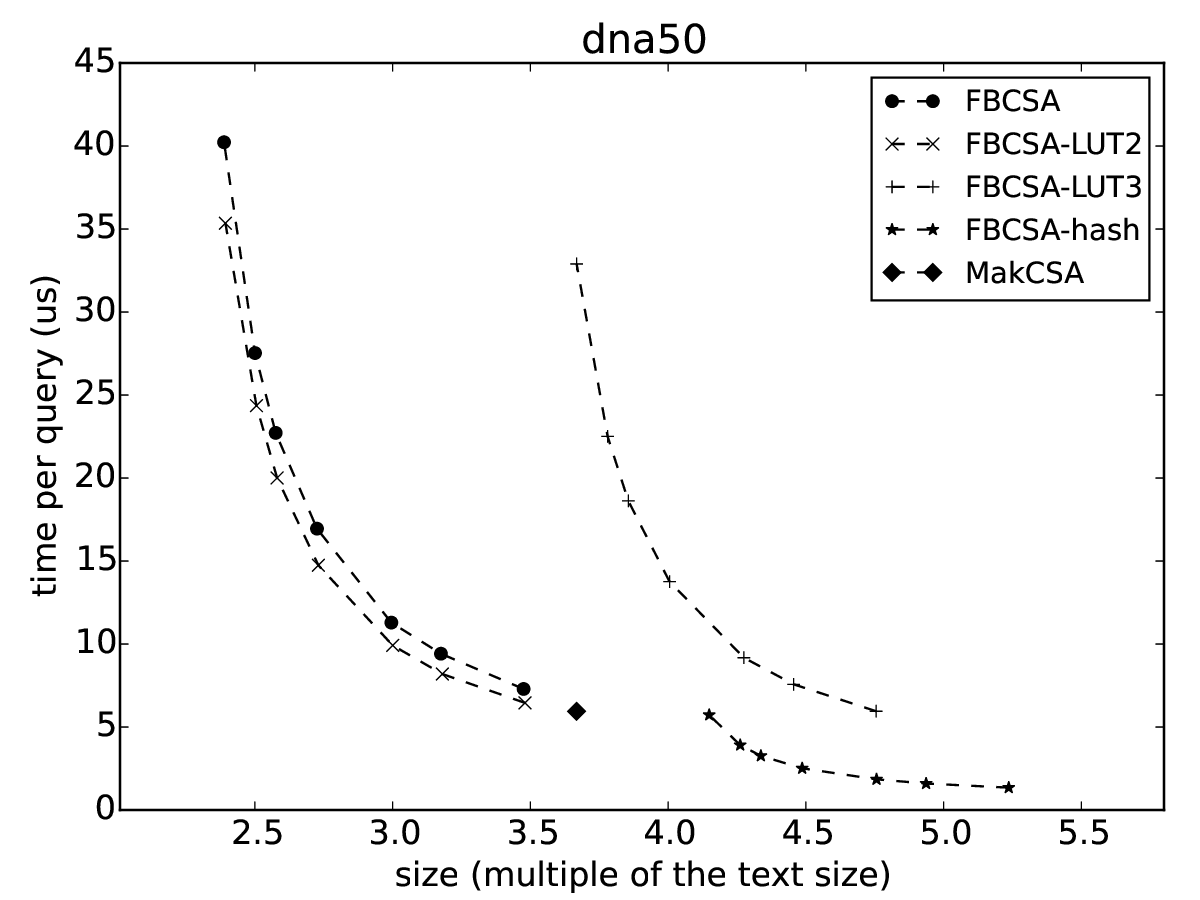}
\includegraphics[width=0.49\textwidth,scale=1.0]{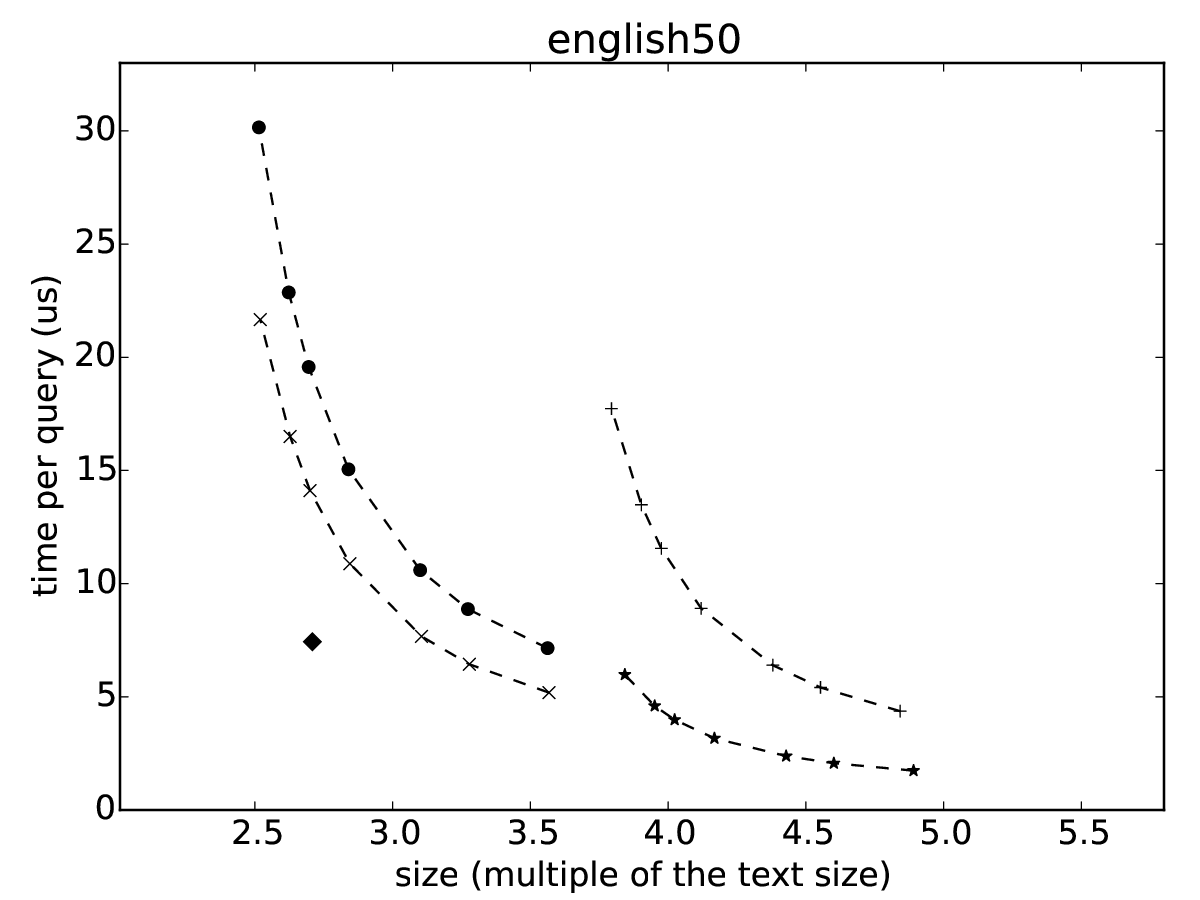}
}
\centerline{
\includegraphics[width=0.49\textwidth,scale=1.0]{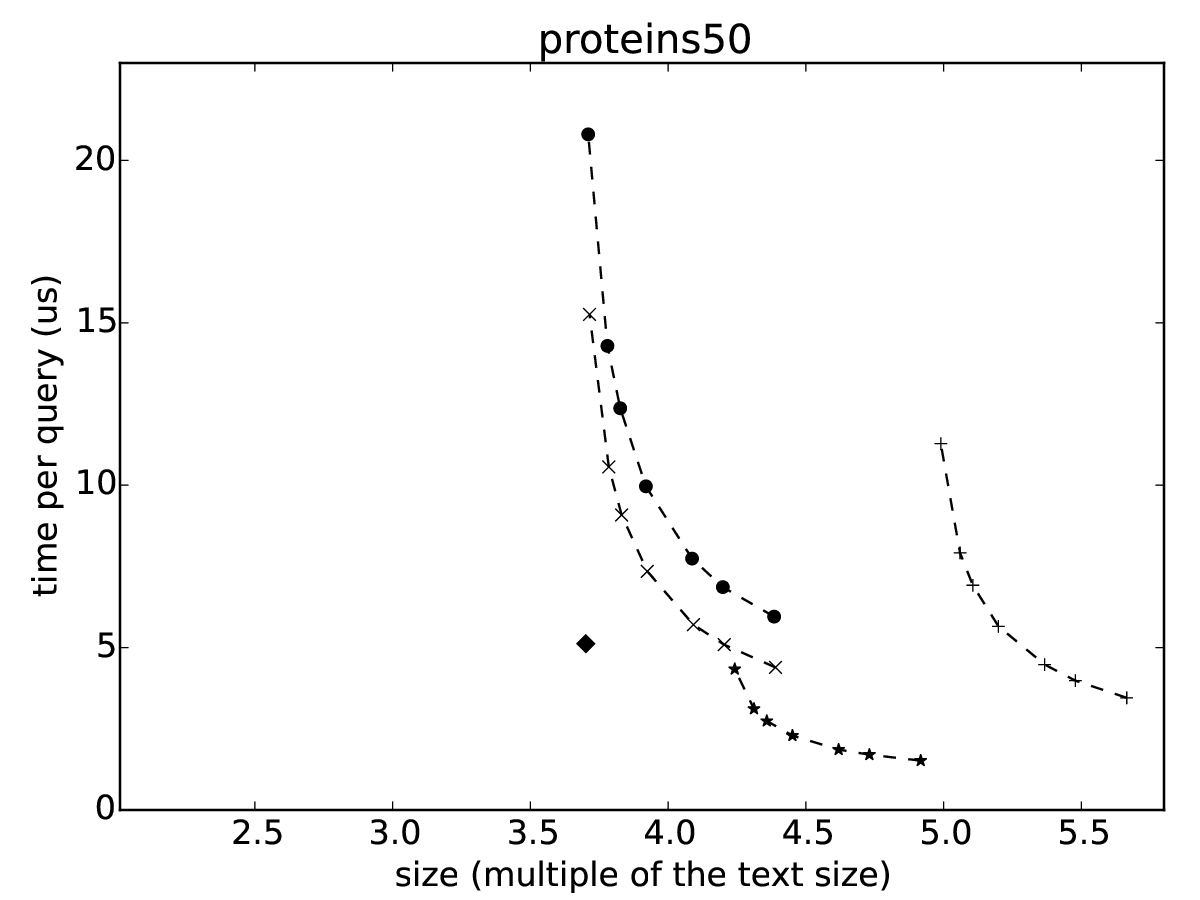}
\includegraphics[width=0.49\textwidth,scale=1.0]{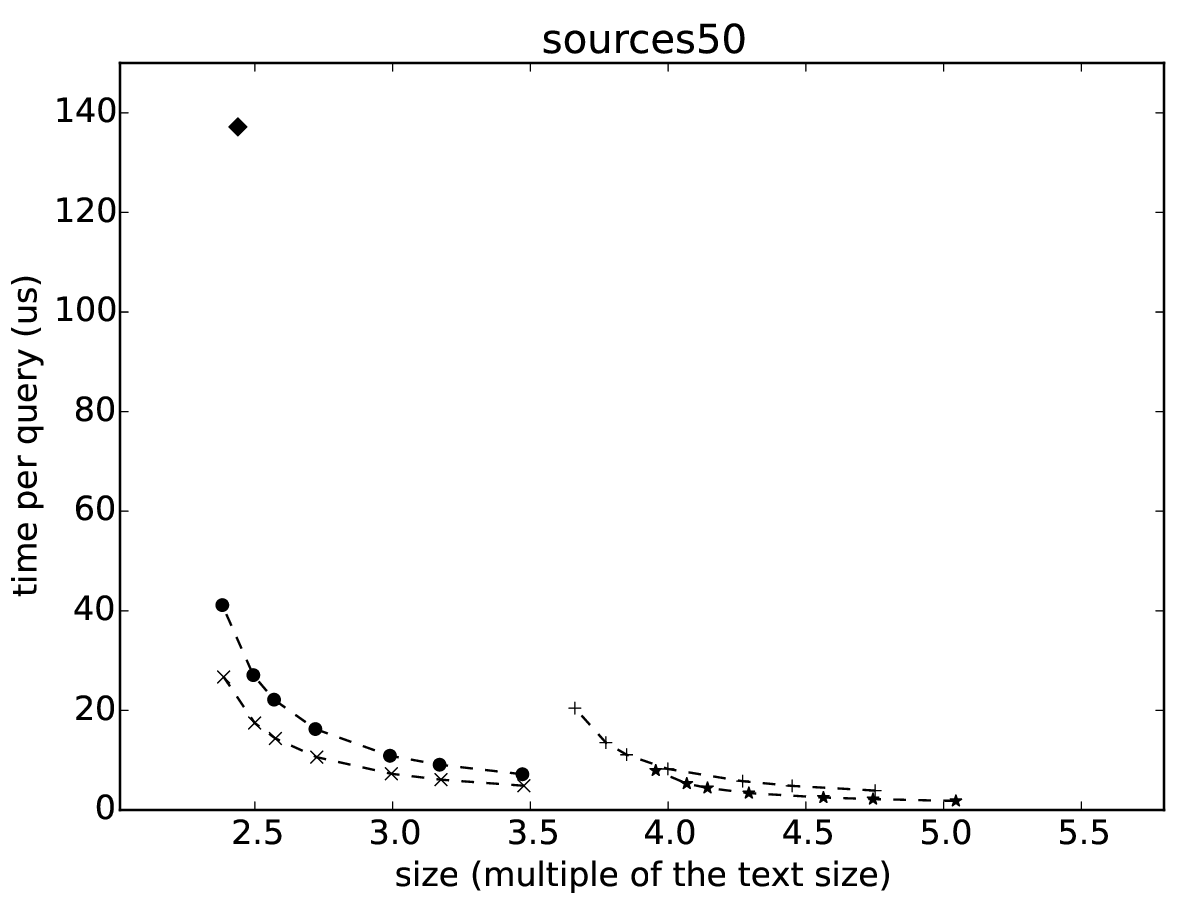}
}
\centerline{
\includegraphics[width=0.49\textwidth,scale=1.0]{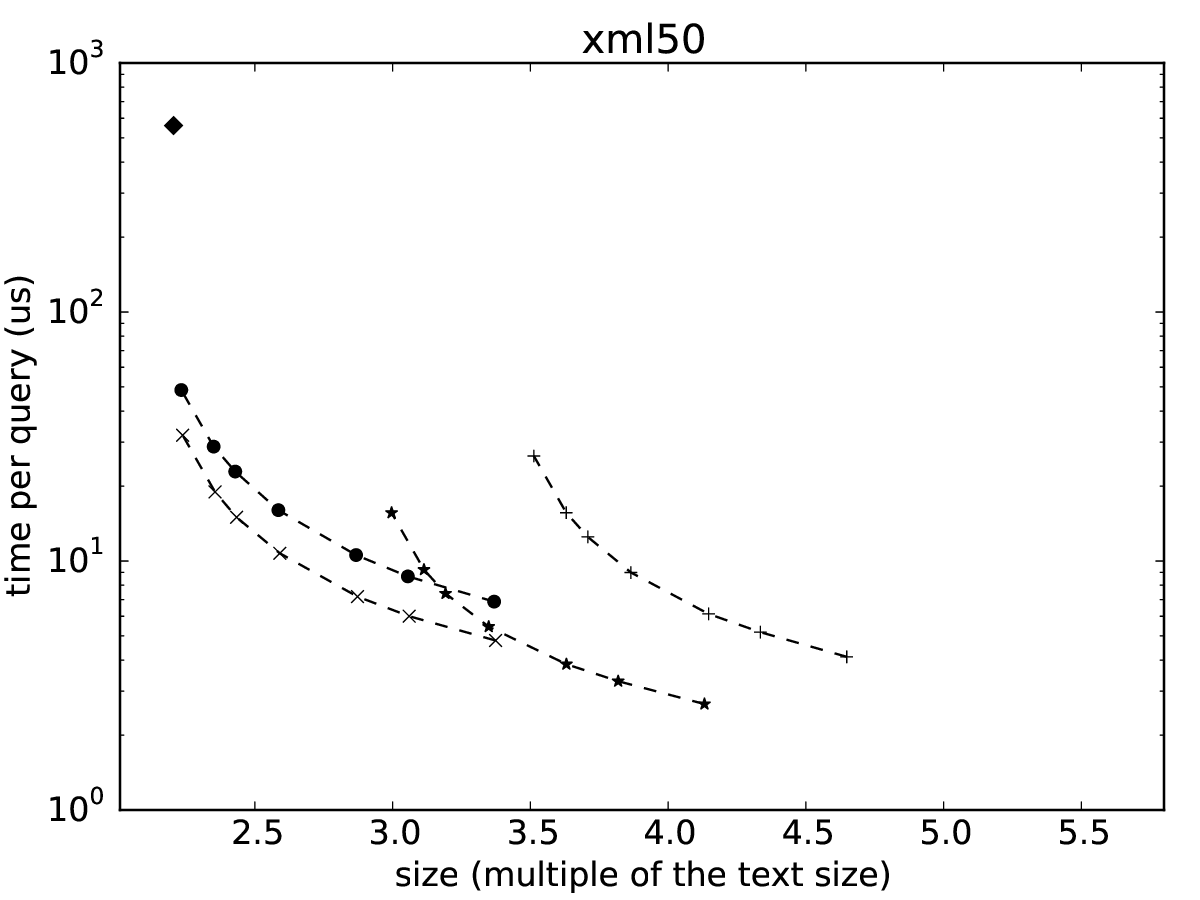}
}
\caption[Results]
{Pattern search time (count query) for FBCSA variants ($bs = 32$) and 
M{\"a}kinen's CSA.
The different results in a series are obtained from varying the sampling 
parameter $ss$ in $\{3, 4, 5, 8, 12, 16, 32\}$.
All times are averages over 500K random patterns of the same length 
$m = 16$.
The patterns were extracted from the respective texts.
Note the logarithmic scale for the \texttt{xml50} dataset.}
\label{fig:fb_mak}
\end{figure}
\begin{figure}
\centerline{
\includegraphics[width=0.49\textwidth,scale=1.0]{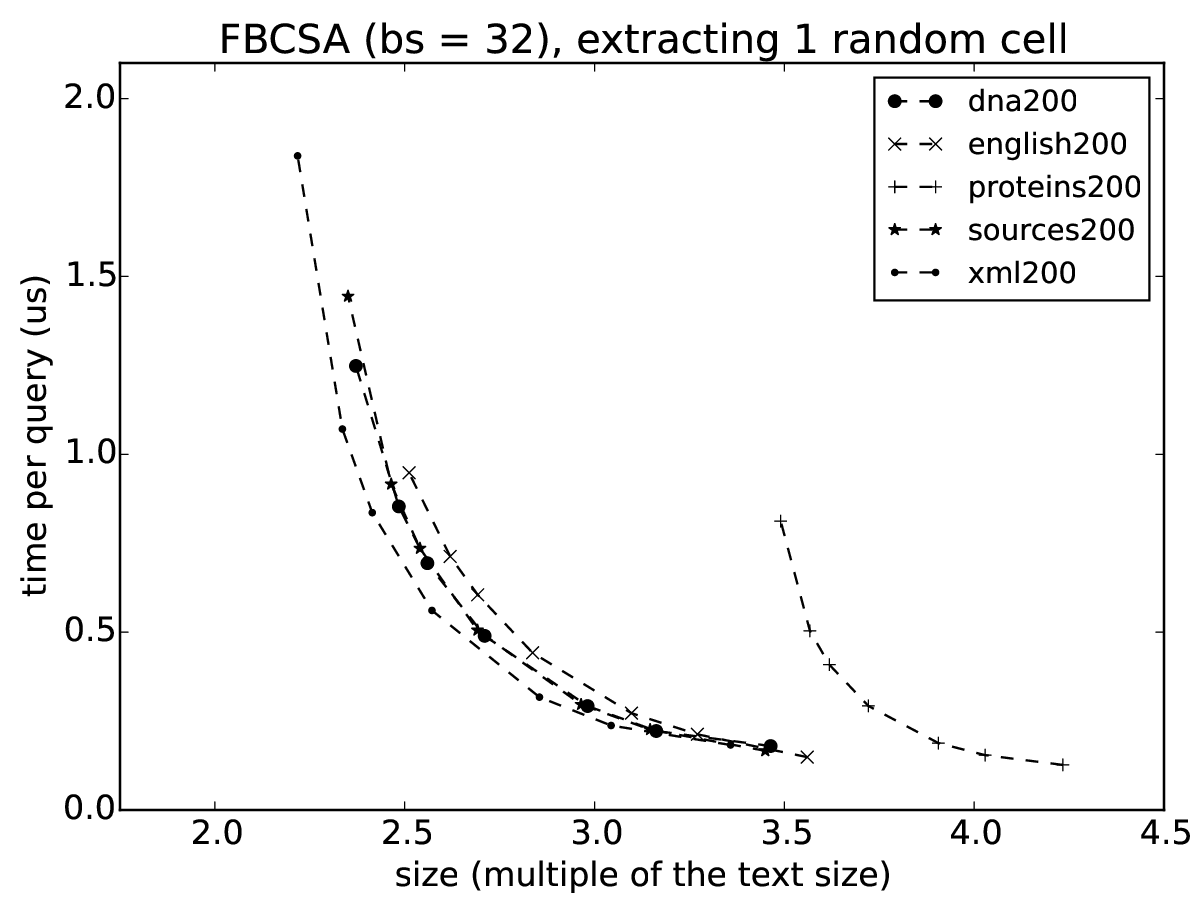}
\includegraphics[width=0.49\textwidth,scale=1.0]{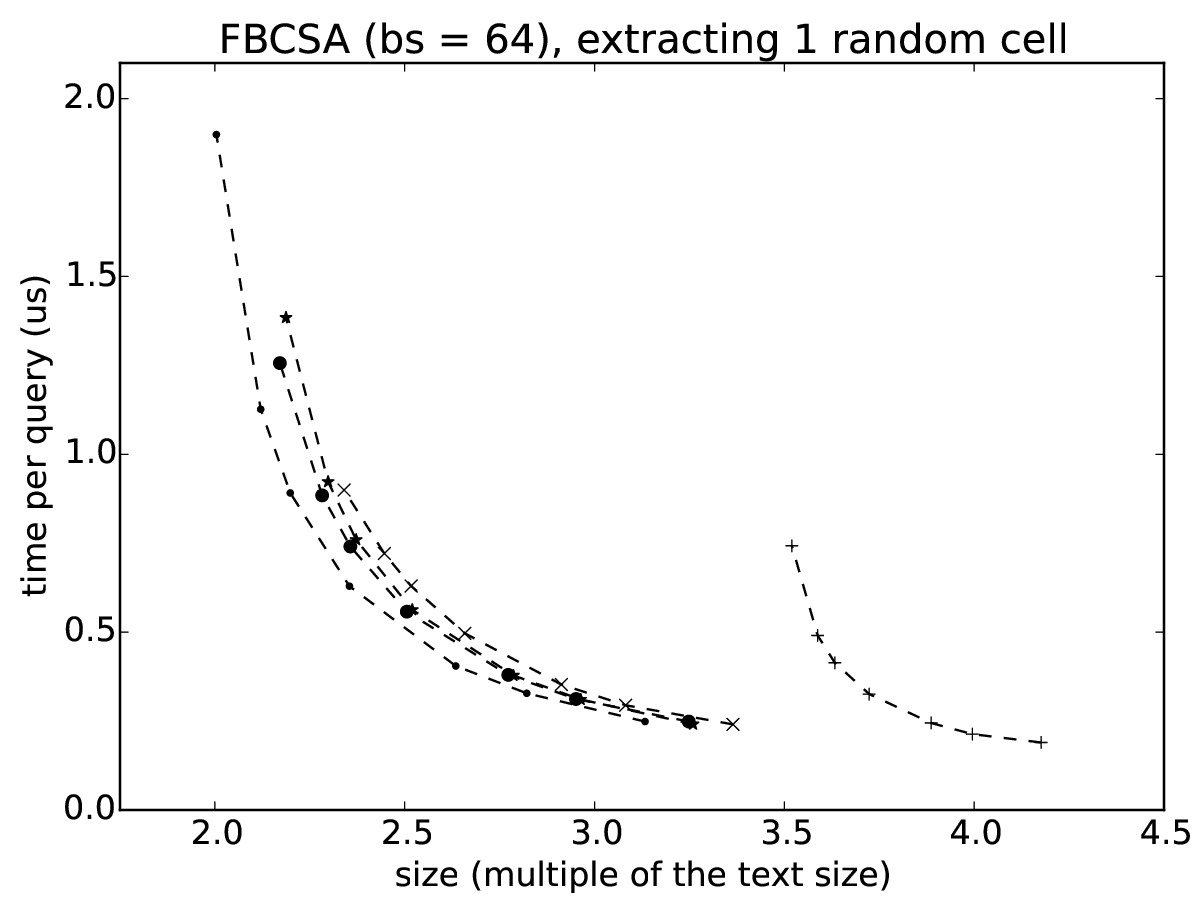}
}
\caption[Results]
{FBCSA index sizes and cell access times with varying $ss$ parameter 
($\{3, 4, 5, 8, 12, 16, 32\}$).
The parameter $bs$ was set to 32 (left figure) or 64 (right figure).
The times are averages over 10M random cell accesses.}
\label{fig:times2}
\end{figure}
\begin{figure}
\centerline{
\includegraphics[width=0.49\textwidth,scale=1.0]{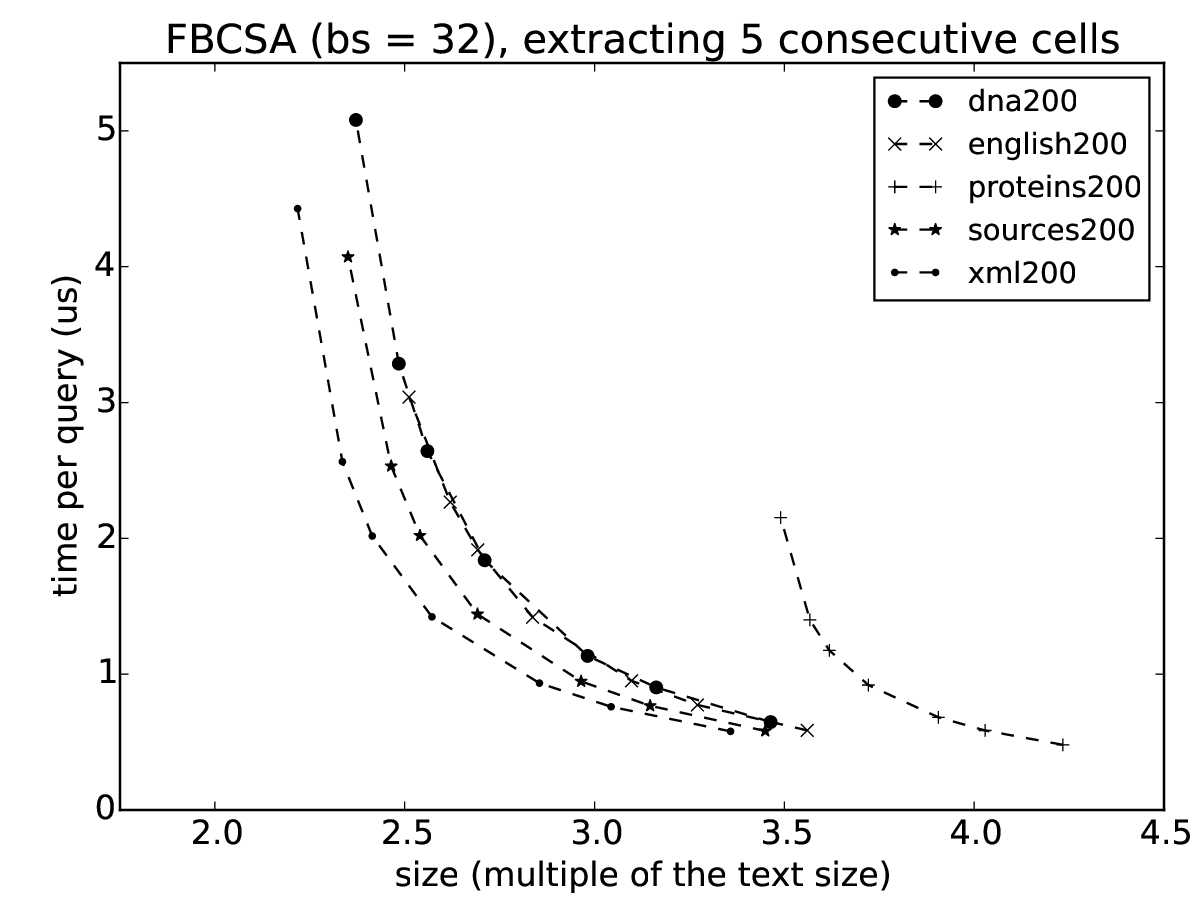}
\includegraphics[width=0.49\textwidth,scale=1.0]{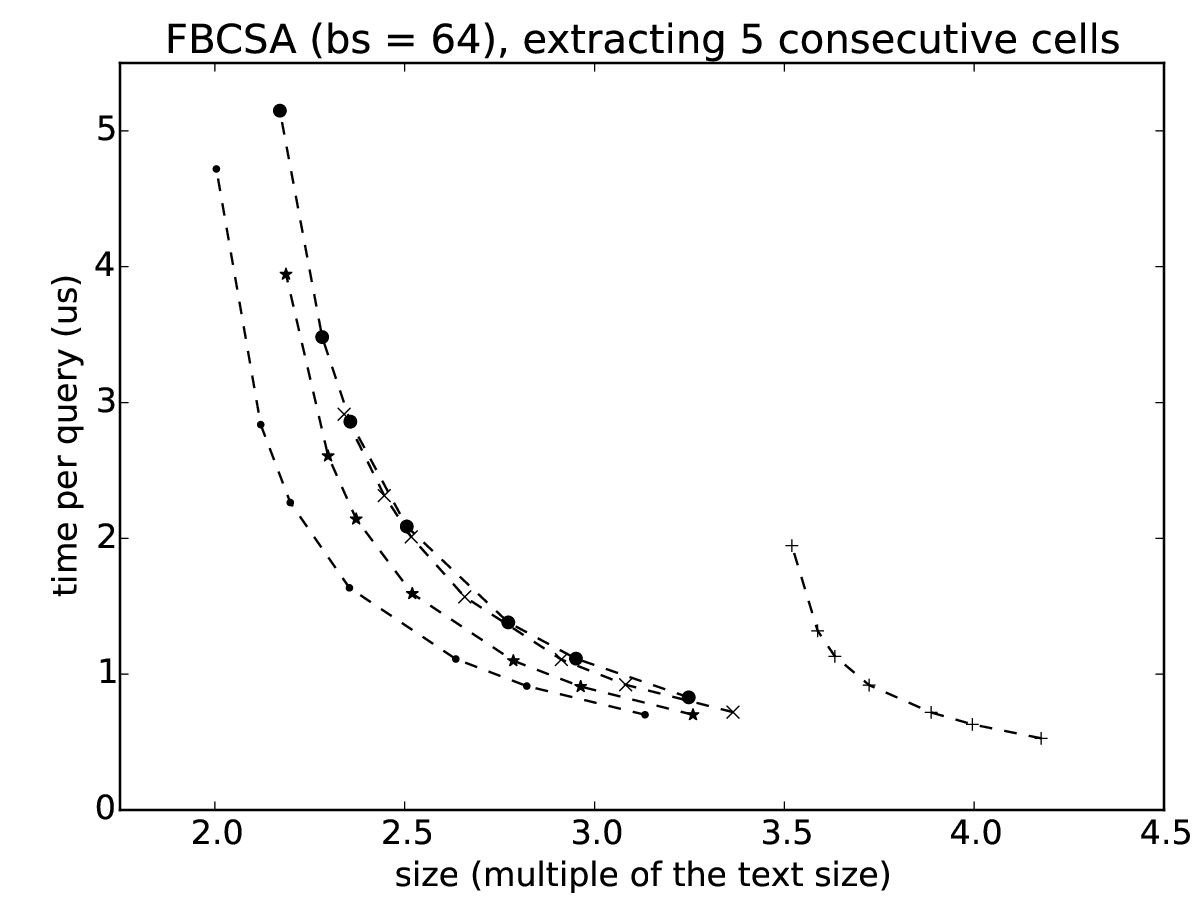}
}
\centerline{
\includegraphics[width=0.49\textwidth,scale=1.0]{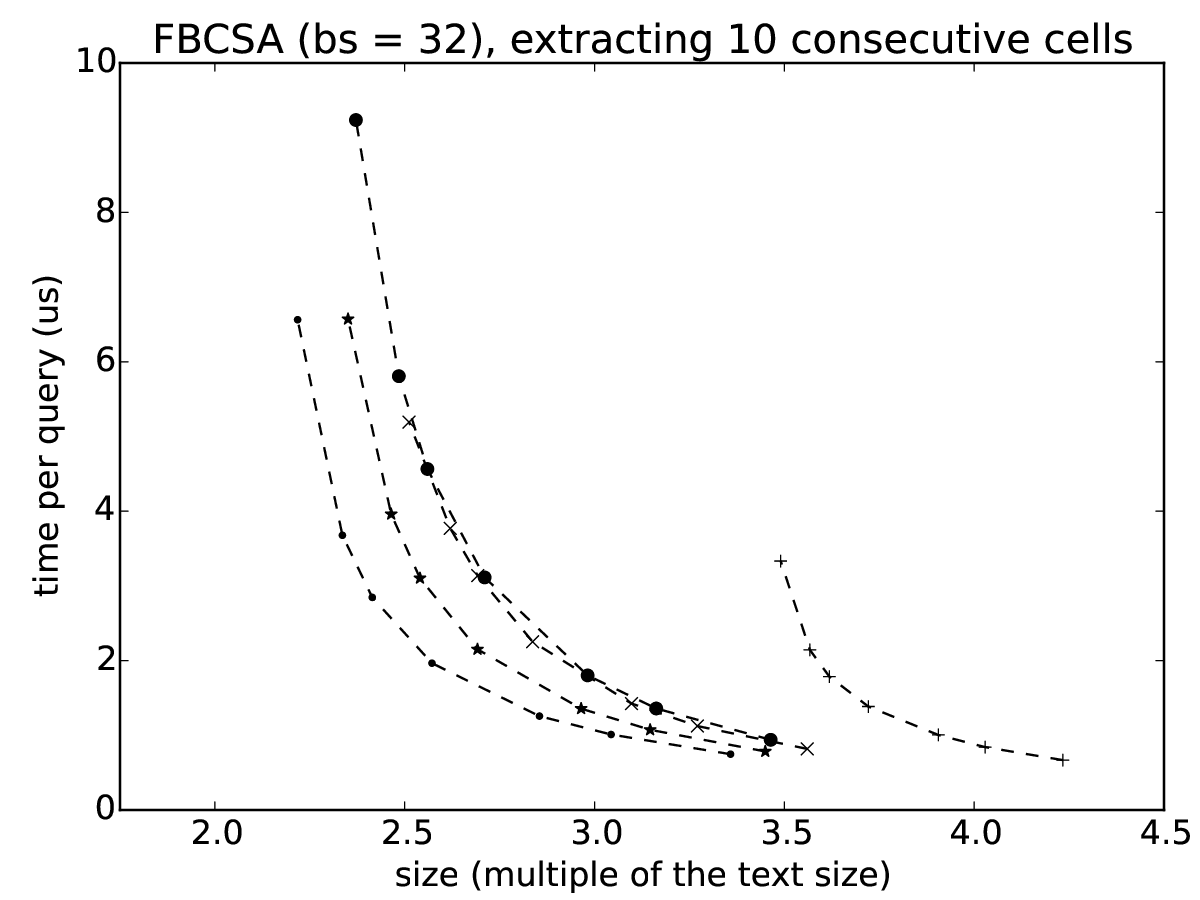}
\includegraphics[width=0.49\textwidth,scale=1.0]{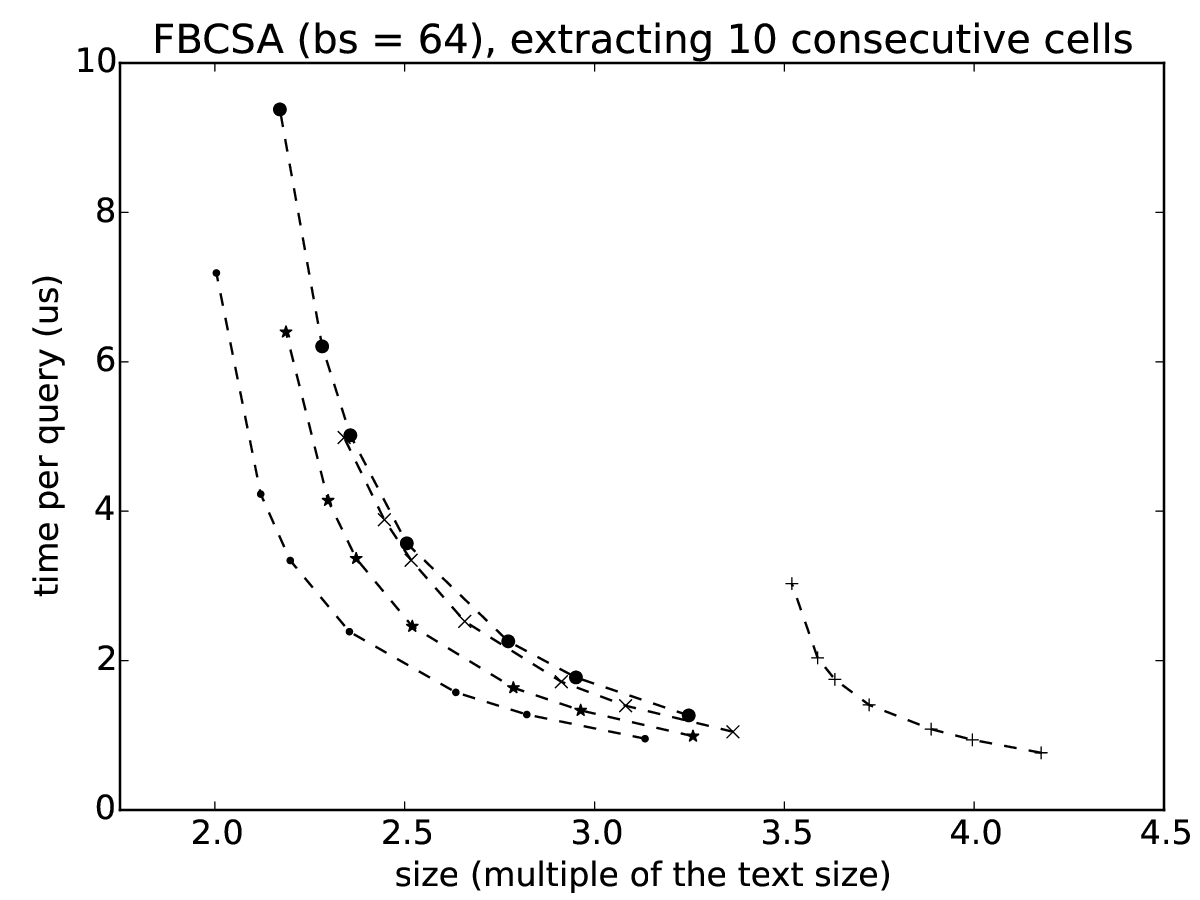}
}
\caption[Results]
{FBCSA, extraction time for $c = 5$ (top figures) and $c = 10$ (bottom figures) 
consecutive cells, with varying $ss$ parameter 
($\{3, 4, 5, 8, 12, 16, 32\}$).
The parameter $bs$ was set to 32 (left figures) or 64 (right figures).
The times are averages over 1M random cell run extractions.}
\label{fig:times3}
\end{figure}

We also compared FBCSA variants against MakCSA in search (count) queries.
Alas, it was possible to use MakCSA only for 50-megabyte datasets.
The results of our comparison are shown in Fig.~\ref{fig:fb_mak}.
MakCSA wins on \texttt{proteins50} and \texttt{english50}, 
is comparable to our variants on \texttt{dna50}, and 
clearly loses on \texttt{sources50} and \texttt{xml50} 
(note the logarithmic scale for the last dataset).
Also, we can add two remarks.
First, the relative overhead of the lookup tables (LUT2 and LUT3) is roughly 
4 times smaller for 200-megabyte datasets, yet (as mentioned) 
MakCSA does not support such large datasets.
Second, the hash component of the index may be optimized for the FBCSA indexes, 
with hopefully more competitive space-time tradeoffs.

\section{Conclusions}

We presented two simple full-text indexes.
One, called SA-hash, speeds up standard suffix array searches with 
reducing significantly the initial search range, thanks to a hash table 
storing range boundaries of all intervals sharing a prefix of a specified 
length.
Despite its simplicity, we are not aware of such use of hashing in 
exact pattern matching, and the approximately 3-fold speedups compared 
to a standard SA may be worth the extra space in many applications.

The other presented data structure is a compact variant of the suffix array, 
related to M{\"a}kinen's compact SA~\cite{DBLP:journals/fuin/Makinen03}.
Our solution works on blocks of fixed size, which provides int32 alignment 
of the layout.
This index is rather fast in single cell access, but not competitive 
if many (e.g., 100) consecutive cells are to be extracted.

Several aspects of the presented indexes require further study.
In the SA-hash scheme collisions in the HT may be eliminated 
with 
perfect hashing.
This should also reduce the overall space use.
In case of plain text, the standard suffix array component may be replaced 
with a suffix array on words~\cite{DBLP:conf/cpm/FerraginaF07}, with possibly 
new interesting space-time tradeoffs.
The idea of deep buckets may be incorporated into some compressed indexes, 
e.g., to save on the several first LF-mapping steps in the FM-index.

\section{Acknowledgement}

The work was supported by the Polish National Science Centre under the project DEC-2013/09/B/ST6/03117 (both authors).

\label{lastpage}
\end{document}